\documentclass[reprint,twocolumn,amsmath,amssymb,aps,pra,longbibliography,floatfix]{revtex4-1}
\usepackage{physics}
\usepackage{gensymb}
\usepackage{amsmath}
\usepackage{amsthm} 
\usepackage{cases}
\usepackage{url}
\usepackage{natbib}
\usepackage{textcase}
\usepackage{amssymb}
\usepackage{graphicx}
\usepackage{bbm} 
\usepackage{calligra} 

\usepackage{mathrsfs}
\DeclareMathAlphabet{\mathpzc}{OT1}{pzc}{m}{it} 
\usepackage{blkarray} 
\usepackage{xparse} 
\NewDocumentCommand{\tens}{e{_^}}{%
  \mathbin{\mathop{\otimes}\displaylimits
    \IfValueT{#1}{_{#1}}
    \IfValueT{#2}{^{#2}}
  }%
}
\usepackage{nicematrix}
\usepackage{bm} 
\usepackage{times}
\usepackage{float}
\usepackage{multirow,microtype,color,relsize}
\usepackage{subfigure}
\usepackage[breaklinks=true]{hyperref}
\usepackage[utf8]{inputenc}
\usepackage[english]{babel}
\hypersetup{colorlinks=true,linkcolor=blue,citecolor=blue}
\hypersetup{linktocpage}
\usepackage{CJKutf8}
\usepackage{breakcites}
\usepackage{float}
\usepackage[dvipsnames]{xcolor}
\definecolor{mypine}{RGB}{1, 121, 111}

\usepackage{simpler-wick} 
\usepackage{MnSymbol}

\begin{document}
\title{Perturbative renormalization group approach to magic-angle twisted bilayer graphene using topological heavy fermion model}
\author{Yi Huang}
\email{yihphysics@gmail.com}
\author{Yang-Zhi Chou}
\email{yzchou@umd.edu}
\author{Sankar Das Sarma}
\affiliation{Condensed Matter Theory Center and Joint Quantum Institute, Department of Physics, University of Maryland, College Park, Maryland 20742, USA}

\begin{abstract}	
We develop a perturbative renormalization group (RG) theory for the topological heavy fermion (THF) model, describing magic-angle twisted bilayer graphene (MATBG) as an emergent Anderson lattice. Our theory focuses on an energy window where the interactions can be treated perturbatively within the THF model, providing insights into the low-energy physics. In particular, the realistic parameters place MATBG near an intermediate regime where the Hubbard interaction $U$ and the hybridization energy $\gamma$ are comparable, motivating the need for RG analysis. Our approach analytically tracks the flow of single-particle parameters and Coulomb interactions within an energy window below $0.1$ eV, providing implications for distinguishing between Kondo-like ($U\gg \gamma$) and projected-limit/Mott-semimetal ($U\ll \gamma$) scenarios at low energies.
We show that the RG flows generically lower the ratio $U/\gamma$ and drive MATBG toward the chiral limit, consistent with the previous numerical study based on the Bistritzer-MacDonald model.
The framework presented here also applies to other moiré systems and stoichiometric materials that admit a THF description, including magic-angle twisted trilayer graphene, twisted checkerboard model, and Lieb lattice, among others, providing a foundation for developing low-energy effective theories relevant to a broad class of topological flat-band materials.

\end{abstract}

\maketitle

\section{Introduction}
Magic-angle twisted bilayer graphene (MATBG) is a prime platform for studying strongly correlated physics in a topologically nontrivial band structure. 
Its nearly flat moiré bands at the magic angle, where the interlayer tunneling is equal to the moiré momentum displacement multiplied by graphene Dirac velocity, lead to interaction-driven phases including correlated insulators and superconductivity~\cite{Cao:2018a,Cao:2018b,Polshyn:2019,Sharpe:2019,Jiang:2019,Lu:2019,Yankowitz:2019,Kerelsky:2019,Xie:2019,Choi:2019,Cao:2020,Serlin:2020,Arora:2020,Wong:2020,Liu:2020,Cao_Nematicity:2021,Choi:2021,Oh:2021,Hao:2021,Zhou:2021a,Zhou:2021b,Park:2021a,Park:2021b,Liu:2021,Kuiri:2022,Su:2023,Tanaka:2025,Pixley:2019,Andrei:2020,Balents:2020,Zondiner:2020,Andrei:2021,Nuckolls:2023}. 
Meanwhile, electronically modeling the flat band in MATBG has faced a fundamental challenge: an isolated flat band in MATBG cannot be described by a tight-binding model due to its topological character and long-range (power-law) Wannier tails.
Two notable categories of approaches have been proposed to resolve the Wannier obstruction: earlier methods that include many remote bands, and the recently introduced topological heavy fermion (THF) model~\cite{Song:2022,H_Shi:2022,Calugaru:2023}. 
The first approach is to combine the flat bands with a few remote bands calculated from Bistritzer-MacDonald (BM) model~\cite{BM_model:2011} such that the whole set of bands is topologically trivial~\cite{Zou:2018,Po:2019,Carr:2019,Song:2021}.
The second approach is the THF model~\cite{Song:2022,H_Shi:2022,Calugaru:2023}, which maps the low-energy physics onto a Anderson–lattice-like problem~\cite{Anderson:1961,Anderson:1970} of $f$-orbitals localized in moiré lattice hybridizing with dispersive $c$-electrons, making it natural to apply heavy-fermion physics tools (e.g., dynamical mean-field theory) to MATBG~\cite{haule:2019,Bascones:2020,Datta:2023,Rai:2024}. 
The THF model provides a conceptually clear division: the localized $f$-fermions (flat-band states with completely quenched kinetic energy, centered on AA sites) carry Hubbard-like Coulomb repulsion, while the itinerant c-fermions (Dirac-like extended states carrying nontrivial Berry curvature) carry the band topology.
The topological flat bands emerge from a band inversion (hybridization) between the $f$- and $c$-orbitals.
Unlike the first approach, which incorporates a few remote bands to resolve the Wannier obstruction, the THF model achieves this by selectively hybridizing only part of the remote-band states near the $\Gamma$ point with the flat bands.
This yields a minimal, analytically tractable, and physically motivated basis of $c$- and $f$-fermions, capturing the essential topological and interaction-driven physics without involving the full remote-band manifold~\cite{Herzog-Arbeitman:2024}.

\begin{figure}[t]
    \centering
    \includegraphics[width=0.9\linewidth]{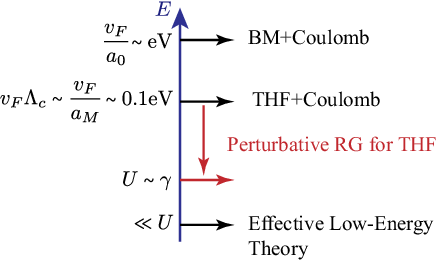}
    \caption{Energy scales of MATBG. $v_F/a_0$ is the energy scale below which the graphene dispersion is approximated by the Dirac cone, where $a_0$ is the graphene lattice constant. $v_F/a_M$ characterizes the kinetic energy associated with the moir\'e lattice of period $a_M$. $\gamma$ denotes the bandgap separating the flat bands from remote bands. $U$ represents the Coulomb interaction strength between electrons in the flat bands.}
    \label{fig:RG_scheme}
\end{figure}

Compared with the numerically intensive BM model, the THF model, parameterized by fitting only a few key quantities—such as the $c$-fermion velocity $v_{\star}$, $cf$-hybridization strength $\gamma$, and flat-band width $M$—offers significant analytical advantages for understanding MATBG’s low-energy properties below $v_{\star}\Lambda_c \simeq 100$ meV.
See Fig.~\ref{fig:RG_scheme}.
$\Lambda_c$ is the scale set by the first moiré Brillouin zone (mBZ) and $v_{\star} \Lambda_c$ defines the bandwidth of the $c$-fermions. 
Specifically, the THF model enables analytical derivations of the flat-band wavefunctions and associated Berry curvature~\cite{Song:2022,Hu:2025}. 
Furthermore, projecting the Coulomb interaction onto the THF single-particle basis naturally disentangles the interactions into clearly defined physical channels. 
These include Hubbard-like onsite repulsion ($U$) among $f$-fermions, density-density interactions within itinerant $c$-fermions ($V$), density-density interactions between localized and itinerant fermions ($W$), exchange coupling ($J$), and double-hybridization processes ($J_{+}$)~\cite{Song:2022}. This decomposition facilitates a transparent interpretation and systematic exploration of correlation effects in MATBG.

\begin{figure}[t]
    \centering
    \includegraphics[width=\linewidth]{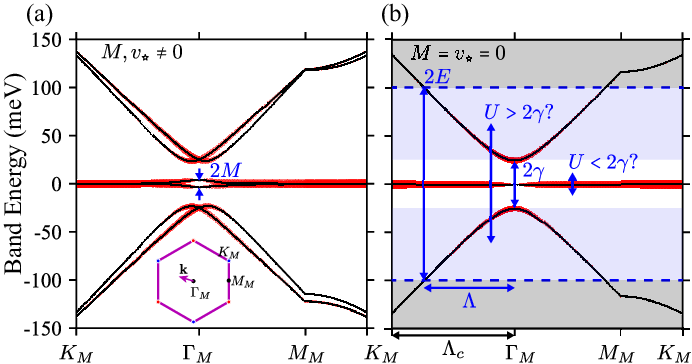}
    \caption{Band structures of the single-particle THF model fitted from the BM model with twist angle $\theta = 1.05^\circ$ and interlayer tunneling ratio $w_0/w_1 = 0.8$. The model parameters are $v_\star = 4.303$ eV·Å and $\gamma = 24.75$ meV~\cite{Song:2022}. Size of red dots represents the $f$-orbital weight.
    (a) Parameters: $M = 3.697$ meV and $v_{\smallstar} = 1.623$ eV·Å.
    (b) $M = v_{\smallstar} = 0$. 
    High-energy modes above the energy cutoff $E$ (dark shaded region) are integrated out. In our perturbative RG scheme, we continue lowering $E$ until it reaches the bottom of the remote bands (light shaded region), where the running cutoff $\Lambda$ flows from $\Lambda_c$ to 0.}
    \label{fig:energy_hierarchy}
\end{figure}

An essential question in interacting MATBG is the relative energy scale hierarchy between $U$, $\gamma$, and $M$. Given that $\gamma$ characterizes the gap between flat and remote bands in the THF model, the existence of flat bands naturally implies $\gamma \gg M$. 
See Fig.~\ref{fig:energy_hierarchy} (a). 
We focus on the strongly correlated regime of the flat band, while the dispersive remote bands remain in the perturbative regime, characterized by the hierarchy $v_{\star} \Lambda_c \gg U \gg M$. Within this regime, two distinct scenarios emerge depending on the relative strength of $U$ and $\gamma$: either $U \gg \gamma$ or $U \ll \gamma$.
See Fig.~\ref{fig:energy_hierarchy} (b). 
In the first scenario, $U \gg \gamma$, the low-energy physics near integer filling can be effectively described by a Kondo lattice model, where spin-spin interactions among local moments dominate while effectively freezing out charge fluctuations ~\cite{Chou:2023a,Hu:2023a,Hu:2023b,Zhou:2024,Chou:2023b}.
This scenario closely resembles the conventional Anderson lattice model, where large $U$ strongly suppresses charge fluctuations, leaving primarily spin degrees of freedom~\cite{Coleman_2015}. 
Between integer fillings, it could be described by a mixed-valence model, analogous to Coulomb blockade physics in quantum dots, where gate voltage tuning induces significant charge fluctuations by bringing adjacent valence states closer in energy~\cite{Li:2023,Lau:2023,zhao2025mixedvalencemottinsulator}. 
In contrast, the second scenario, $U \ll \gamma$, allows projection of remote bands out, leaving isolated flat bands that form a Mott semimetal at the charge neutrality point (CNP). 
Here, interactions split the initially degenerate flat bands by an energy scale of $U$ away from the $\Gamma_M$ point, while preserving a Dirac-like band touching at $\Gamma_M$ due to the underlying topological structure that couples the spin and charge degrees of freedom~\cite{Ledwith:2024,Hu:2025,Zhao:2025,Mott_gap}.

Practically, however, realistic ab initio calculations~\cite{Song:2022,Calugaru:2023,Rai:2024} and experimental observations~\cite{Wong:2020,Choi:2021,Oh:2021,Nuckolls:2023} suggest that with a lattice dielectric constant $\kappa=6$ and screening from double gates at a distance $d\simeq10$ nm, the Coulomb energy scale $U\simeq40$ meV is comparable to $\gamma\simeq30$ meV. Such an intermediate regime ($U\sim\gamma$) presents considerable theoretical challenges, as it bridges the distinct conceptual limits of the $U \gg \gamma$ and $U \ll \gamma$ scenarios. Consequently, theoretical analyses in this intermediate regime become significantly more intricate, making it challenging to clearly identify the dominant physical picture. Performing a systematic renormalization group (RG) analysis on the parameter flow of $U/\gamma$ within the THF model is thus highly beneficial. 

In this paper, we develop a general perturbative RG framework based on the THF model to systematically analyze the flow of single-particle parameters ($v_{\star}, \gamma$) and interaction energies ($U, V, W, J, J_{+}$) within the energy window $U < |E| < v_{\star} \Lambda_c \simeq 0.1$ eV. 
As shown in Fig.~\ref{fig:RG_scheme}, this energy window allows for a perturbative treatment, despite the fact that the Coulomb interaction (set by $U$) is greater than the mini-band bandwidth.
This analysis qualitatively and quantitatively predicts how these parameters evolve. 
Given the practical regime $U \sim \gamma$, our RG approach provides valuable insights into the energy hierarchy between $U$ and $\gamma$ as well as the resultant low-energy physics at energies below $U$.

Early pioneering RG work by Vafek and Kang (VK)~\cite{Vafek:2020} introduces a two-stage RG scheme that systematically integrates high-energy degrees of freedom in the BM model and calculates how the parameters in the BM model such as the interlayer tunnelings $w_0/w_1$ in the AA/AB region and graphene Dirac velocity $v_F$ flow toward low energy.
Similar renormalization ideas have also been applied recently in a self-consistent Hartree–Fock study of a tight-binding model~\cite{Miguel:2025}.
The two-stage RG are described by two energy windows, $w_{1}<\abs{E}<v_F/a_0$ and $ \abs{E} <w_{1}$, where $w_{1} \sim 100$ meV, $v_F/a_0 \sim 1$ eV is the graphene Dirac cone band width, and $a_0 = 0.246$ nm is the graphene lattice constant.
VK showed that Coulomb interaction suppresses the ratio $w_0/w_1$ and drives the system toward the chiral limit ($w_0/w_1\to 0$) as the energy cutoff is lowered. Meanwhile, both $w_1$ and $v_F$ increase proportionally, preserving the magic-angle condition $w_1/(v_F \Lambda_c) \simeq 1$, where the Dirac cone displacement is given by $\Lambda_c\sim 4\pi/3 a_M$ and $a_M=a_0/[2 \sin (\theta/2)]$ is the moir\'e lattice constant~\cite{BM_model:2011}.
Since $v_{\star}$ is of the same order as $v_F$, the energy window $|E| < w_1$ in VK’s second-stage RG naturally coincides with the energy window $|E| < v_{\star} \Lambda_c$ in our THF-based formalism.

Unlike VK’s non-perturbative treatment of the second-stage RG--accessible only numerically--our RG analysis using the THF model remains perturbative in this same energy window and yields analytical results.
Our RG results indicate that $v_{\star}$ and $\gamma$ both increase by roughly a factor of $1.3$, whereas $U$ remains nearly unchanged as the energy scale is progressively reduced toward $U$ and remote bands are integrated out (cf. Fig.~\ref{fig:gamma_vstar_RG}). 
By comparing THF model parameters fitted from the BM model with varying $w_0/w_1$ ratios~\cite{Song:2022,Calugaru:2023}, we find an increase in both $v_{\star}$ and $\gamma$ as the system moves toward the chiral limit ($w_0/w_1$ decreases), consistent with VK's findings. 
However, VK did not explicitly discuss the hierarchy between $U$ and $\gamma$. 
Our RG scheme based on the THF model fills this gap by demonstrating that the ratio $U/\gamma$ decreases as the energy cutoff is lowered. 
Consequently, even if the initial condition at $|E| = v_{\star}\Lambda_c$ satisfies $U>\gamma$, the projected limit $U\ll \gamma$ near CNP can be approached under the condition $v_{\star}\Lambda_c/U \gg 1$.
Additionally, we compute RG flows for coupling constants associated with other interaction channels. Our analysis shows that the interaction parameters $V$ and $W$ exhibit minor changes (less than 10\%), whereas the exchange interaction $J$ increases and the double-hybridization interaction $J_{+}$ decreases, resulting in an approximately twofold increase in the ratio $J/J_{+}$. These RG flows are summarized comprehensively in Fig.~\ref{fig:interaction_flow}.

The structure of the paper is as follows.
Section~\ref{sec:RG_scheme} presents the perturbative RG formalism based on the THF model, where we derive the one-loop diagrams that govern the flow equations.
Section~\ref{sec:main_results} provides the RG analysis and the corresponding main results, including the detailed derivation of the flow equations.
We conclude in Section~\ref{sec:conclusion} by summarizing our findings
and discuss remaining open questions. 
Appendix~\ref{sec:Green_function_identities} summarizes key integrals and identities for readers interested in technical details, and Appendix~\ref{sec:RG_beyond_leading} presents all one-loop diagrams beyond the leading order and the corresponding RG equations for completeness.

\section{Model and formalism}
\label{sec:RG_scheme}
\begin{figure}
    \centering
    \includegraphics[width=0.9\linewidth]{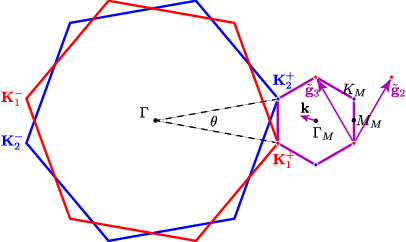}
    \caption{Brillouin zone of twisted bilayer graphene. Two graphene layers are labeled by 1 (red) and 2 (blue) with a twist $\theta$. The moir\'e Brillouin zone (purple) is shown near the graphene $\vb*{K}^+$ valleys, with moir\'e lattice vectors $\tilde{\vb{g}}_2$ and $\tilde{\vb{g}}_3$.}
    \label{fig:TBG_BZ}
\end{figure}
In this section, we describe our perturbative RG scheme using the THF model.
Our goal is to establish how the single-particle parameters ($v_{\star}, \gamma$) and interaction strengths ($U, V, W, J, J_+$) evolve under the RG flow as high-energy remote-band degrees of freedom are integrated out.
To achieve this goal, we formulate the action $S[c, f]$ including the interaction terms, and perform a momentum-shell RG by integrating out fast modes in a thin energy shell near the cutoff.

\subsection{THF model}
We begin with the full Hamiltonian of the THF model, which describes localized $f$-orbitals hybridizing with dispersive $c$-orbitals in the moiré lattice of MATBG \cite{Song:2022}. 
\begin{align}
    &H_0 = - \mu_f \sum_{\eta,s} \sum_{\vb{R}} f^{\dagger}_{\vb{R},\alpha,\eta,s} f_{\vb{R},\alpha,\eta,s} + \nonumber \\
    &+\sum_{\eta s} \sum_{a,a'} \sum_{\abs{\vb{k}} < \Lambda} [h_{a,a'}^{(c)\eta}(\vb{k}) - \mu_c \delta_{a,a'}] c^{\dagger}_{\vb{k},a,\eta,s} c_{\vb{k},a',\eta,s} + \nonumber\\
    &+ \frac{1}{\sqrt{N_M}} \sum_{\eta,s} \sum_{\abs{\vb{k}} < \Lambda} \sum_{\vb{R}} \sum_{\alpha,a} [e^{ i \vb{k} \cdot \vb{R}} h^{(fc)\eta}_{\alpha,a} (\vb{k}) f^{\dagger}_{\vb{R},\alpha,\eta,s} c_{\vb{k},a,\eta,s}  + h.c.]
\end{align}
where $a=1,2,3,4$ labels the $c$-fermions and $\alpha=1,2$ labels the $f$-fermions. 
$\eta=\pm1$ labels the graphene $\vb{K}^{\pm}$ valley (cf. Fig.~\ref{fig:TBG_BZ}), and $s=\pm1$ labels the spin.
The Hamiltonian is defined with a momentum cutoff $\Lambda$ that includes remote $c$-states beyond the first mBZ~\cite{c_band_cutoff}, as shown in Fig.~\ref{fig:TBG_BZ}. 
$N_M$ is the total number of moir\'e unit cells in the finite system, which equals the number of $\vb{k}$ points sampled in the first mBZ. $\vb{R}$ denotes the real-space center of the $f$ orbital in each cell.
We adopt the Fourier transform of $f$-orbitals as follows
\begin{align}
    f_{\vb{R},\alpha,\eta,s} = \frac{1}{\sqrt{N_M}} \sum_{\vb{k} \in \mathrm{mBZ}} e^{i \vb{k} \cdot \vb{R}} f_{\vb{k},\alpha,\eta,s}, \\
    f_{\vb{k},\alpha,\eta,s} = \frac{1}{\sqrt{N_M}} \sum_{\vb{R}} e^{-i \vb{k} \cdot \vb{R}} f_{\vb{R},\alpha,\eta,s}. \label{eq:f_k_FT}
\end{align}
If $\vb{k}$ of an $f$-fermion is outside the first mBZ in Eq.~\eqref{eq:f_k_FT}, it should be folded back to the first mBZ denoted as $(\vb{k}/{\tilde{\vb{g}}}) \in $ mBZ so that $\vb{k} = m_1 \tilde{\vb{g}}_1 + m_2 \tilde{\vb{g}}_2 + \vb{k}/{\tilde{\vb{g}}}$, where $\tilde{\vb{g}}$ are the moiré reciprocal lattice vectors shown in Fig.~\ref{fig:TBG_BZ}.
As a result, the non-interacting Hamiltonian can be rewritten as 
\begin{align}
    &H_0 = - \mu_f \sum_{\eta,s} \sum_{\vb{k} \in \mathrm{mBZ}} f^{\dagger}_{\vb{k},\alpha,\eta,s} f_{\vb{k},\alpha,\eta,s} + \nonumber\\
    &+\sum_{\eta s} \sum_{a,a'} \sum_{\abs{\vb{k}} < \Lambda} [h_{a,a'}^{(c)\eta}(\vb{k}) - \mu_c \delta_{a,a'}] c^{\dagger}_{\vb{k},a,\eta,s} c_{\vb{k},a',\eta,s} + \nonumber\\
    &+ \sum_{\eta,s} \sum_{\abs{\vb{k}} < \Lambda} \sum_{\alpha,a} [h^{(fc)\eta}_{\alpha,a} (\vb{k}) f^{\dagger}_{\vb{k}/{\tilde{\vb{g}}},\alpha,\eta,s} c_{\vb{k},a,\eta,s}  + h.c.],
\end{align}
where the last term may describe an umklapp process that couples momentum in different mBZs when $\Lambda$ is large compared to the first mBZ.
Here, since we focus on the low-energy physics below the first remote bands, we set $\Lambda=\Lambda_c$ and $\Lambda_c=K_M\Gamma_M=4\pi/3 a_M$ as the first mBZ boundary (cf. Fig.~\ref{fig:TBG_BZ}), and we ignore the umklapp process coming from higher remote bands so that $\vb{k}/\tilde{\vb{g}}=\vb{k}$~\cite{mBZ}.
To simplify the problem further, we approximate the hexagonal first mBZ by a circle of radius $\Lambda_c$, so that $\vb{k} \in \mathrm{mBZ} \to \abs{\vb{k}} < \Lambda_c$. 
In this way, the original THF model Hamiltonian is diagonal in the momentum sector $H_0=\sum_{\eta,s}\sum_{\abs{\vb{k}}<\Lambda_c} \Psi^{\eta s\dagger}_{\vb{k}} h_0^{\eta}(\vb{k}) \Psi^{\eta s}_{\vb{k}}$
by choosing the basis as $\Psi=\{f_1,f_2,c_1,c_2,c_3,c_4\}$, where we drop the valley/spin and momentum indices for simplicity.
The dispersion can be solved by diagonalizing the $6\times6$ Hamiltonian
\begin{align}
    h_0^{\eta}(\vb{k}) =
    \begin{bmatrix}
        h^{(f)\eta}(\vb{k}) - \mu_f & h^{(fc)\eta}(\vb{k}) \\
        h^{(cf)\eta}(\vb{k}) & h^{(c)\eta}(\vb{k}) - \mu_c
    \end{bmatrix},
\end{align}
where $h^{(f)\eta} = 0_{2\times 2}$,
\begin{align}
    h^{(c)\eta} &= 
    \begin{bmatrix}
        0_{2\times 2} & v_{\star}(\eta k_x \sigma_0 + i k_y \sigma_z) \\
        v_{\star}(\eta k_x \sigma_0 - i k_y \sigma_z) & M \sigma_x
    \end{bmatrix}, 
\end{align}
and 
\begin{align}
    h^{(fc)\eta}  = h^{(cf)\eta\dagger} =
    \begin{bmatrix}
        \gamma \sigma_0 + v_{\smallstar}(\eta k_x \sigma_x + k_y \sigma_y), 0_{2\times 2}
    \end{bmatrix}.
\end{align}
In the expression of $h^{(fc)\eta}$, $\gamma$ encodes the momentum-independent hybridization and the $v_{\smallstar}$ term encodes the momentum-dependent hybridization \cite{Song:2022}.

To proceed analytically, we adopt the approximation $M = 0$, $v_{\smallstar} = 0$~\cite{Song:2022}, justified by the hierarchy $M, v_{\smallstar} \Lambda_c \ll v_{\star} \Lambda_c$. Setting $M=0$ and $v_\smallstar=0$ allows for analytical expressions of the remote-band Green functions, and we stop our theory at an energy scale much larger than $M$ (e.g., at the scale of $U$, about ten times larger than $M$). The proper effective low-energy theory for $E\ll U$ should adapt the renormalized parameters (based on our theory) and restore $M$, but we focus only on the renormalization in the energy window $v_\star\Lambda_c>E>U$ in this work. Strictly speaking, $|v_\star/v_\smallstar|\approx2.65$ does not justify setting $v_\smallstar=0$. However, neglecting $v_\smallstar$ is unlikely to induce a significant difference in our theory, as the main effects due to $v_\smallstar$ are splitting the remote bands and shifting the $k$ points of the remote-band band edge. In addition, the ratio $|v_\star/v_\smallstar|$ becomes larger as the system approaches the chiral limit (i.e., decreasing $w_0/w_1$ in the BM model)~\cite{Song:2022}. Our result and the earlier VK result \cite{Vafek:2020} both indicate that MATBG flows toward the chiral limit under RG. Thus, the approximation $v_\smallstar=0$ becomes progressively more reliable under RG flows.
We should also emphasize that this approximation $M = 0$, $v_{\smallstar} = 0$ in the THF model is different from the ``chiral limit'' in the BM model where $w_0/w_1=0$. The bare parameters of the THF model at $\Lambda_c$ are extracted using $w_0/w_1=0.8$.
This simplification allows us to diagonalize $H_0$ analytically and project the system into flat and remote band subspaces. 

Under the simplification, $M = 0$ and $v_{\smallstar} = 0$, the non-interacting Hamiltonian $H_0$ decouples into two independent $3 \times 3$ blocks labeled by the orbital index $\alpha = 1,2$~\cite{Hu:2025}.
The two decoupled sectors of the Hamiltonian formed by the basis $\Phi_{\alpha}=\{f_{\alpha}, c_{\alpha}, c_{\alpha+2}\}$, $\alpha=1,2$ are given by
\begin{align}
    h_0^{\eta}(\vb{k}) = h_0^{\alpha=1,\eta} \oplus h_0^{\alpha=2,\eta},
\end{align}
where
\begin{align}
    h_0^{\eta\alpha} &= 
    \begin{bmatrix}
        0 & \gamma & 0 \\
        \gamma & 0 & v_{\star} (\eta k_x + is_{\alpha} k_y) \\
        0 & v_{\star} (\eta k_x - is_{\alpha} k_y) & 0
    \end{bmatrix}, \\
    &=
    \begin{bmatrix}
        0 & \gamma & 0 \\
        \gamma & 0 & v_{\star} \abs{\vb{k}} \eta e^{is_{\alpha} \eta \theta_{\vb{k}}} \\
        0 & v_{\star} \abs{\vb{k}} \eta e^{-is_{\alpha} \eta \theta_{\vb{k}}} & 0
    \end{bmatrix}.
\end{align}
Here, $s_{\alpha}=(-1)^{\alpha+1}$ and $\theta_{\vb{k}}$ is the angle of $\vb{k}$ relative to ${\hat k}_x$ axis.
For each $\alpha$ sector, the eigenenergies read
\begin{align}
    E_0=0, E_{\pm} = \pm \sqrt{v_{\star}^2 \abs{\vb{k}}^2 + \gamma^2}, 
\end{align}
and the corresponding eigenstates are given by
\begin{align}
    \ket{\psi_{0,\vb{k}}^{\alpha \eta}} &= \frac{[v_{\star} \abs{\vb{k}} \eta e^{is_{\alpha} \eta \theta_{\vb{k}}}, 0, -\gamma]^{\mathrm{T}}}{\sqrt{v_{\star}^2 \abs{\vb{k}}^2 + \gamma^2}}, \label{eq:psi_0}\\
    \ket{\psi_{\pm,\vb{k}}^{\alpha \eta}} &= \frac{[\gamma \eta e^{is_{\alpha} \eta \theta_{\vb{k}}}, \pm \eta e^{is_{\alpha} \eta \theta_{\vb{k}}}\sqrt{v_{\star}^2 \abs{\vb{k}}^2 + \gamma^2} , v_{\star} \abs{\vb{k}}]^{\mathrm{T}}}{\sqrt{2(v_{\star}^2 \abs{\vb{k}}^2 + \gamma^2)}}.\label{eq:psi_pm}
\end{align}
The Berry curvature and the resulting Chern numbers for the flat band eigenstates Eq.~\eqref{eq:psi_0} are derived explicitly, revealing a $U(4) \times U(4)$ symmetry due to the decomposition into $C = \pm 1$ Chern sectors.
The Berry curvature reads
\begin{align}
    \Omega_{\vb{k}}^{\alpha\eta} &= \hat z \cdot [\grad_{\vb{k}} \cross \bra{\psi_{0,\vb{k}}^{\alpha \eta}} i \grad_{\vb{k}} \ket{\psi_{0,\vb{k}}^{\alpha \eta}}], \\
    &= -\frac{2v_{\star}^2 \gamma^2 \eta s_{\alpha}}{(v_{\star}^2 \abs{\vb{k}}^2 + \gamma^2)^2},
\end{align}
which gives the Chern number for each flat band 
\begin{align}
    C = \frac{1}{2\pi} \int d^2k \; \Omega_{\vb{k}}^{\alpha\eta} = -\eta s_{\alpha}.
\end{align}
We see that the whole flat-band manifold has in total 8 bands labeled by $\alpha$, $\eta$, and $s$, which can be decomposed into $C=+1$ and $C=-1$ Chern sectors where each of them has 4 bands that can be related by a $U(4)$ rotation.
As a result, $H_0$ has $U(4)\times U(4)$ symmetry in the limit of $M \to 0$, $v_{\smallstar} \to 0$~\cite{mass}.

\begin{figure}
    \centering
    \includegraphics[width=0.8\linewidth]{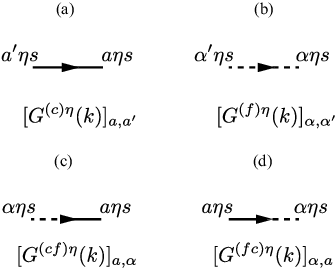}
    \caption{(a) $c$-fermion propagator. (b) $f$-fermion propagator. (c-d) Anomalous $cf$-propagator. The solid line represents $c$-fermion, and the dashed line represents $f$-fermion.}
    \label{fig:propagator}
\end{figure}
\begin{figure*}[t]
    \centering
    \includegraphics[width=\linewidth]{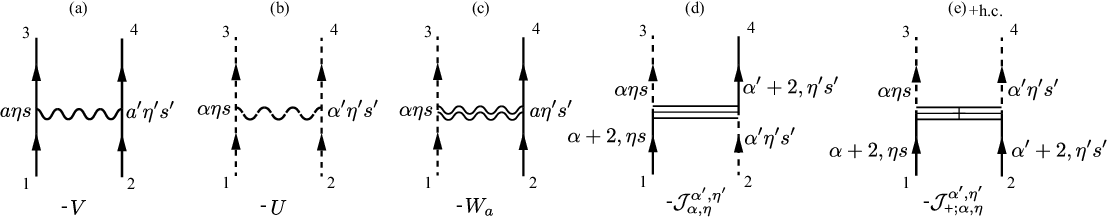}
    \caption{Coulomb interactions between $c$- and $f$-fermions. $c$-fermion ($f$-fermion) is denoted as a solid (dashed) line with an arrow. Due to symmetry, $W_1=W_2$, $W_3=W_4$. $\mathcal{J}_{\alpha,\eta}^{\alpha',\eta'} = J(\eta\eta' + s_{\alpha} s_{\alpha'})/2$ and $\mathcal{J}_{+;\alpha,\eta}^{\alpha',\eta'} = -J_+(\eta\eta' -  s_{\alpha} s_{\alpha'})/2$ where $s_{\alpha}=(-)^{\alpha-1}$.}
    \label{fig:Coulomb_interactions}
\end{figure*}

\subsection{RG formalism}
We construct the imaginary-time action of the THF model defined with a momentum cutoff $\Lambda$ 
\begin{align}
    S[c,f] = S_0[c,f] + S_I[c,f],
\end{align}
where $S_0$ is non-interacting action and $S_I$ represents the interactions.
The partition function is given by
\begin{align}
    Z = \int \mathcal{D}[c,f] e^{-S[c,f]}.
\end{align}
We adopt the perturbative momentum-shell RG scheme such that we integrate out the high-energy $c$- and $f$-fermion modes in the remote bands to analyze the effective action in the low-energy regime (cf. Fig.~\ref{fig:energy_hierarchy} b).
Focusing on CNP where $\mu_c=\mu_f=0$ and using the approximation $M=v_{\smallstar}=0$, $S_0$ is given by
\begin{align}
    S_0[c,f] = \int_k {\bar \Phi}_{k}^{\alpha\eta s} (-i\omega \mathbbm{1} + h_0^{\alpha\eta}(\vb{k})) \Phi_{k}^{\alpha\eta s},
\end{align}
where we use Einstein's summation convention to sum over the dummy indices and integrate over $k=\{\omega,\vb{k}\}$, the Matsubara frequency $\omega$ and momentum $\vb{k}$,
\begin{align}
    \int_k = \int_{-\infty}^{\infty} \frac{d\omega}{2\pi} \int \frac{d^2\vb{k}}{(2\pi)^2} \Theta(\Lambda -\abs{\vb{k}}).
\end{align}
The band basis $\{{\bar \Phi}_{k}^{\alpha \eta s}\}$ is defined as
\begin{align}
    \Phi_{k}^{\alpha\eta s} &= [f_{k,\alpha,\eta, s}, c_{k,\alpha, \eta, s},c_{k,\alpha+2, \eta, s}]^{\mathrm{T}}, \\
    \bar \Phi_{k}^{\alpha\eta s} &= [{\bar f}_{k,\alpha,\eta, s}, {\bar c}_{k,\alpha, \eta, s}, {\bar c}_{k,\alpha+2, \eta, s}].
\end{align}
The non-interacting action can be projected onto the flat bands and remote bands by a change of basis $\{{\bar \psi}_{b,k}^{\alpha \eta s} = {\bar \Phi}_{k}^{\alpha \eta s} \ket{\psi_{b,\vb{k}}^{\alpha\eta}}\}$ where $b=0,\pm$,
\begin{align}
    S_0[c,f] =S_{0,0}[\psi_0] + S_{0,1}[\psi_{\pm}],
\end{align} 
where the projected actions read
\begin{align}
    S_{0,0}&= \int_k {\bar \Phi}_{k}^{\alpha\eta s} P_0 (-i\omega \mathbbm{1} + h_0^{\alpha\eta}(\vb{k})) P_0 \Phi_{k}^{\alpha\eta s}, \nonumber \\
    &=\int_k {\bar \psi}_{0,k}^{\alpha \eta s} (-i\omega \mathbbm{1} + E_0) \psi_{0,k}^{\alpha \eta s} , \\
    S_{0,1}&= \int_k {\bar \Phi}_{k}^{\alpha\eta s} P_1 (-i\omega \mathbbm{1} + h_0^{\alpha\eta}(\vb{k})) P_1 \Phi_{k}^{\alpha\eta s}, \nonumber \\
    &= \int_k \sum_{b=\pm}{\bar \psi}_{b,k}^{\alpha \eta s} (-i\omega \mathbbm{1} + E_b(\vb{k})) \psi_{b,k}^{\alpha \eta s}.
\end{align}
Here, $P_0 = \ket{\psi_{0,\vb{k}}^{\alpha \eta}}\bra{\psi_{0,\vb{k}}^{\alpha \eta}}$ and $P_1=\sum_{b=\pm}\ket{\psi_{b,\vb{k}}^{\alpha \eta}}\bra{\psi_{b,\vb{k}}^{\alpha \eta}} = \mathbbm{1}- P_0$ are the corresponding projection operators to the flat bands and remote bands, respectively.
The non-interacting Green function $G^{\alpha \eta}(k) = (i\omega \mathbbm{1} - h_0^{\alpha\eta}(\vb{k}))^{-1}$ reads 
\begin{align}
    &G^{\alpha \eta} (z=i\omega, \vb{k}) = \sum_{b=0,\pm} \frac{\ket{\psi_{b,\vb{k}}^{\alpha \eta}}\bra{\psi_{b,\vb{k}}^{\alpha \eta}}}{z-E_b}  , \nonumber\\
    &= G^{\alpha \eta}_0 (z, \vb{k}) + G^{\alpha \eta}_1 (z, \vb{k}),\label{eq:Green_function}
\end{align}
where $G_0$ and $G_1$ are Green functions projected to the flat bands and the remote bands, respectively:
\begin{align}
    &G^{\alpha \eta}_0 (z=i\omega, \vb{k}) = \frac{\ket{\psi_{0,\vb{k}}^{\alpha \eta}}\bra{\psi_{0,\vb{k}}^{\alpha \eta}}}{z -E_0} = \frac{1}{z(v_{\star}^2 \abs{\vb{k}}^2 + \gamma^2)} \times \nonumber \\
    &\times 
    \begin{bmatrix}
        v_{\star}^2 \abs{\vb{k}}^2 & 0 & -\gamma v_{\star} \abs{\vb{k}}\eta e^{is_{\alpha} \eta \theta_{\vb{k}}} \\[1mm]
        0 & 0 & 0 \\[1mm]
        -\gamma v_{\star} \abs{\vb{k}}\eta e^{-is_{\alpha} \eta \theta_{\vb{k}}} & 0 & \gamma^2
    \end{bmatrix},\label{eq:G_0}
\end{align}
\begin{align}
    &G^{\alpha \eta}_1 (z=i\omega, \vb{k}) = \sum_{b=\pm} \frac{\ket{\psi_{b,\vb{k}}^{\alpha \eta}}\bra{\psi_{b,\vb{k}}^{\alpha \eta}}}{z-E_b} = \frac{-1}{-z^2 + v_{\star}^2 \abs{\vb{k}}^2 + \gamma^2} \times \nonumber \\
    &\times
    \begin{bmatrix}
        \frac{z \gamma^2}{v_{\star}^2 \abs{\vb{k}}^2 + \gamma^2} & \gamma & \frac{z \gamma v_{\star} \abs{\vb{k}}\eta e^{is_{\alpha} \eta \theta_{\vb{k}}}}{v_{\star}^2 \abs{\vb{k}}^2 + \gamma^2} \\[1mm]
        \gamma & z & v_{\star} \abs{\vb{k}}\eta e^{is_{\alpha} \eta \theta_{\vb{k}}} \\[1mm]
        \frac{z \gamma v_{\star} \abs{\vb{k}}\eta e^{-is_{\alpha} \eta \theta_{\vb{k}}}}{v_{\star}^2 \abs{\vb{k}}^2 + \gamma^2} & v_{\star} \abs{\vb{k}}\eta e^{-is_{\alpha} \eta \theta_{\vb{k}}} & \frac{z v_{\star}^2 \abs{\vb{k}}^2}{v_{\star}^2 \abs{\vb{k}}^2 + \gamma^2}
    \end{bmatrix}, \nonumber \\
    &= \begin{bmatrix}
        G^{(f)\eta} & G^{(fc)\eta} \\
        G^{(cf)\eta} & G^{(c)\eta}
    \end{bmatrix}.\label{eq:G_1}
\end{align}
Since we only integrate out the remote bands in our RG scheme, we focus on the remote band Green function Eq.~\eqref{eq:G_1} whose diagrams are shown in Fig.~\ref{fig:propagator}.
We divide the remote-band fermion modes into two parts, the slow modes $\psi_{\pm,k}^{<}$ with $\abs{\vb{k}}<\Lambda-d\Lambda$ and the fast modes $\psi_{\pm,k}^{>}$ within a momentum shell $\Lambda-d\Lambda< \abs{\vb{k}}<\Lambda$.
To relate the fermionic operators in the eigenbasis to those in the band basis, we use the notation $\hat O[c_k^{<}, f_k^{<}] \equiv \hat O[\psi_{0,k}, \psi_{\pm,k}^{<}]$ and $\hat O[c_k^{>}, f_k^{>}] \equiv \hat O[\psi_{\pm,k}^{>}]$. 
This notation indicates that these are the same fermionic operators $\hat O$, expressed in different bases.
In particular, the Jacobian of this basis transformation in the path integral equals the identity.
The perturbative RG is performed by integrating the high-energy remote-band fermions $\psi_{\pm}^{>}$, while retaining the flat-band $\psi_{0}$ and low-energy remote-band modes $\psi_{\pm,k}^{<}$. 
The partition function can be rewritten as
\begin{align}
    Z &= \int \mathcal{D}[\psi_0,\psi^{<}_{\pm}] e^{-S_{0,0}[\psi_0]-S_{0,1}[\psi^{<}_{\pm}]} \times\nonumber \\
    &\times\int \mathcal{D}[\psi^{>}_{\pm}]  e^{-S_{0,1}[\psi^{>}_{\pm}]-S_I[c,f]} , \\
    &= \int \mathcal{D}[\psi_0,\psi^{<}_{\pm}] e^{-S_{\mathrm{eff}}[\psi_0,\psi^{<}_{\pm}]},
\end{align}
where the effective action is given by
\begin{align}
    S_{\mathrm{eff}} = S_{0,0}[\psi_0] + S_{0,1}[\psi^{<}_{\pm}] -\ln Z_{0,1}^{>} - \ln[\ev{e^{-S_I}}].
\end{align}
We introduce the partial trace over $\psi^{>}_{\pm}$ fields
\begin{align}
    \ev{\hat O} &= \frac{1}{Z_{0,1}^{>}} \int \mathcal{D}[\psi^{>}_{\pm}] e^{-S_{0,1}[\psi^{>}_{\pm}]} {\hat O}[c,f], \\
    &= \frac{1}{Z_{0,1}^{>}} \int \mathcal{D}[c^{>},f^{>}] e^{-S_{0,1}[c^{>},f^{>}]} {\hat O}[c,f].
\end{align}
Here, $Z_{0,1}^{>}=\int \mathcal{D}[\psi^{>}_{\pm}] e^{-S_{0,1}[\psi^{>}_{\pm}]}$ is the partition function for the fermion modes in the momentum shell of the remote bands, which is a normalization factor of the partial trace.
The non-interacting remote-band Green function Eq.~\eqref{eq:G_1} (cf. Fig.~\ref{fig:propagator}) can be rewritten using this partial trace notation 
\begin{align}
    [G^{(c)\eta}(k)]_{a,a'} &= -\ev{c_{k,a,\eta,s} {\bar c}_{k,a',\eta,s}}, \\
    [G^{(f)\eta}(k)]_{\alpha,\alpha'} &= -\ev{f_{k,\alpha,\eta,s} {\bar f}_{k,\alpha',\eta,s}}, \\
    [G^{(cf)\eta}(k)]_{a,\alpha} &= -\ev{c_{k,a,\eta,s} {\bar f}_{k,\alpha,\eta,s}}, \\
    [G^{(fc)\eta}(k)]_{a,\alpha} &= -\ev{f_{k,\alpha,\eta,s} {\bar c}_{k,a,\eta,s}}.
\end{align}
The Coulomb interaction projected on the band basis reads
\begin{align}\label{eq:S_U}
    &S_U = \frac{U}{2} \int_{\{k_i\}} {\bar f}_{k_3,\alpha,\eta,s}{\bar f}_{k_4,\alpha',\eta',s'} f_{k_2,\alpha',\eta',s'} f_{k_1,\alpha,\eta,s}, \\
    &S_V = \frac{V}{2} \int_{\{k_i\}} {\bar c}_{k_3,a,\eta,s}{\bar c}_{k_4,a',\eta',s} c_{k_2,a',\eta',s} c_{k_1,a,\eta,s}, \\
    &S_W = W_a\int_{\{k_i\}} {\bar f}_{k_3,\alpha,\eta,s} {\bar c}_{k_4,a,\eta',s'} c_{k_2,a,\eta',s'} f_{k_1,\alpha,\eta,s},\\
    &S_J = \mathcal{J}_{\alpha,\eta}^{\alpha',\eta'} \times \nonumber \\
    &\times  \int_{\{k_i\}} {\bar f}_{k_3,\alpha,\eta,s} {\bar c}_{k_4,\alpha'+2,\eta',s'} f_{k_2,\alpha',\eta',s'}  c_{k_1,\alpha+2,\eta,s}, \label{eq:S_J}\\
    &S_{J_{+}} = \frac{1}{2} \mathcal{J}_{+;\alpha,\eta}^{\alpha',\eta'} \times \nonumber \\
    &\times \int_{\{k_i\}} \qty({\bar f}_{k_3,\alpha,\eta,s} {\bar f}_{k_4,\alpha',\eta',s'} c_{k_2,\alpha'+2,\eta',s'} c_{k_1,\alpha+2,\eta,s} + \mathrm{h.c.}),  \label{eq:S_J+}
\end{align}
where $\mathcal{J}_{\alpha,\eta}^{\alpha',\eta'} = J(\eta\eta' + s_{\alpha} s_{\alpha'})/2$ and $\mathcal{J}_{+;\alpha,\eta}^{\alpha',\eta'} = -J_+(\eta\eta' -  s_{\alpha} s_{\alpha'})/2$,
and we use the integral notation
\begin{align}
    &\int_{\{k_i\}} = \int_{\{\abs{\vb{k}_i}<\Lambda\}} \prod_{i=1}^4 \qty[\frac{d^2\vb{k}_i d\omega_i}{(2\pi)^3}] \times \nonumber \\
    &\times {(2\pi)^3} \delta(\vb{k}_1 + \vb{k}_2 - \vb{k}_3 - \vb{k}_4) \delta(\omega_1 + \omega_2 - \omega_3 - \omega_4).
\end{align}
Since the non-interacting action $S_0$ (in the limit of $M,v_{\smallstar} \to 0$) and the interaction terms $S_U,S_V,S_W,S_J,S_{J_{+}}$ all have the $U(4)\times U(4)$ symmetry, the full interacting MATBG system also preserves this symmetry~\cite{Song:2022}.
Here, the matrix elements in $J$ and $J_{+}$ terms enforce certain textures in the valley ($\eta,\eta'$) and orbital ($\alpha,\alpha'$) indices between two interacting electrons. See Table~\ref{tab:J_and_Jplus}.
\begin{table}[t]
    \centering
    \begin{minipage}[t]{0.45\linewidth}
        \centering
        \begin{tabular}{c|c|c}
            $\mathcal{J}_{\alpha,\eta}^{\alpha',\eta'}/J$ & $\eta=\eta'$ & $\eta\neq \eta'$ \\
            \hline
            $\alpha=\alpha'$ & $+$ & $0$ \\
            $\alpha\neq\alpha'$ & $0$ & $-$
        \end{tabular}
    \end{minipage}
    \hspace{0.05\linewidth}
    \begin{minipage}[t]{0.45\linewidth}
        \centering
        \begin{tabular}{c|c|c}
            $\mathcal{J}_{+;\alpha,\eta}^{\alpha',\eta'}/J_{+}$ & $\eta=\eta'$ & $\eta\neq \eta'$ \\
            \hline
            $\alpha=\alpha'$ & $0$ & $+$ \\
            $\alpha\neq\alpha'$ & $-$ & $0$
        \end{tabular}
    \end{minipage}
    \caption{Valley-orbital texture of exchange and double-hybridization interaction. Left: $\mathcal{J}_{\alpha,\eta}^{\alpha',\eta'} = J(\eta\eta' + s_{\alpha} s_{\alpha'})/2$; Right: $\mathcal{J}_{+;\alpha,\eta}^{\alpha',\eta'} = -J_+(\eta\eta' -  s_{\alpha} s_{\alpha'})/2$. Here, $\eta,\eta'\in\{+,-\}$, $\alpha,\alpha'\in\{1,2\}$, and $s_{\alpha}=(-)^{\alpha-1}$.}
    \label{tab:J_and_Jplus}
\end{table}
These interactions are represented diagrammatically in Fig.~\ref{fig:Coulomb_interactions}.
The interaction terms can be evaluated perturbatively through cumulant expansion
\begin{align}\label{eq:cumulant_expansion}
    \ln[\ev{e^{-S_I}}] &= -\ev{S_I} + \frac{1}{2} (\ev{S_I^2}- \ev{S_I}^2) + \dots \nonumber \\
    &+ \frac{(-)^l}{l!} \times \text{$l$-th cumulant of $S_I$} + \dots.
\end{align}
As a result, the one-loop perturbative correction to the interaction can be obtained through
\begin{align}
    \delta S_I = -\ln[\ev{e^{-S_I}}] \approx \ev{S_I} - \frac{1}{2} (\ev{S_I^2}- \ev{S_I}^2).
\end{align}
The first-order term $\delta S_I^{(1)} = \ev{S_I}$ yields the flow equations for $v_{\star}$ and $\gamma$ [cf. the self-energy diagrams shown in Fig.\ref{fig:self_energy_leading} and Eqs.~(\ref{eq:v_star_flow}-\ref{eq:gamma_flow})].
The second-order term $\delta S_I^{(2)} = - \frac{1}{2} (\ev{S_I^2}- \ev{S_I}^2)$ generate the RG flow for $U, V, W, J, J_+$ [cf. the leading one-loop vertex corrections shown in Fig.~\ref{fig:interaction_leading} and Eqs.~(\ref{eq:dV_lead}-\ref{eq:dJp_lead})].

\section{RG analysis and main results}
\label{sec:main_results}
\begin{figure}[t]
    \centering
    \includegraphics[width=0.8\linewidth]{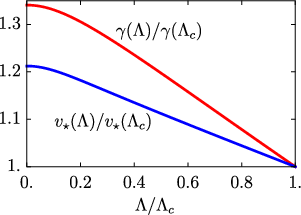}
    \caption{RG flow of $\gamma$ and $v_{\star}$. The initial conditions at $\Lambda_c$ are $v_{\star}=4.303$ eV$\cdot$\AA, $\gamma=24.75$ meV, $V=48.31$ meV, $U=57.95$ meV, $W_1=44.03$ meV, $W_3=50.20$ meV, $J=J_{+}=16.38$ meV.}
    \label{fig:gamma_vstar_RG}
\end{figure}
We now present and derive the main results of our perturbative RG analysis for the THF model of MATBG~\cite{Song:2022}. Our analysis reveals how both single-particle parameters and interaction strengths evolve as high-energy remote-band degrees of freedom are systematically integrated out, providing a self-consistent route to low-energy effective theory near charge neutrality.
The central findings are summarized in Figs.~\ref{fig:gamma_vstar_RG} and \ref{fig:interaction_flow} and Eqs.~(\ref{eq:v_star_flow}-\ref{eq:dJp_lead}), which show the RG flow of the hybridization strength $\gamma$, the $c$-fermion velocity $v_\star$, and various interaction couplings $V$, $U$, $W$, $J$, and $J_{+}$. In the rest of the section, we discuss these main results and provide details of the derivations.

\subsection{RG differential equations}
\label{sec:RG_eq}

\begin{figure}[t]
    \centering
    \includegraphics[width=\linewidth]{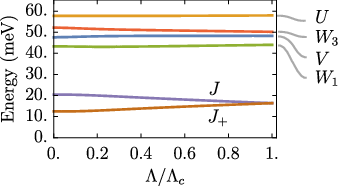}
    \caption{RG flow of interactions. The initial conditions at $\Lambda_c$ are $v_{\star}=4.303$ eV$\cdot$\AA, $\gamma=24.75$ meV, $V=48.31$ meV, $U=57.95$ meV, $W_1=44.03$ meV, $W_3=50.20$ meV, $J=J_{+}=16.38$ meV.}
    \label{fig:interaction_flow}
\end{figure}

With the formalism discussed in the previous section, we analytically derive the leading order one-loop RG equations by integrating out the remote bands.
The one-loop RG equations for $v_\star$ and $\gamma$ are:
\begin{align}\label{eq:v_star_flow}
    \frac{d\ln \abs{v_{\star}}}{d\Lambda}&=-V \frac{1}{8\pi} \frac{(v_{\star}^2 \Lambda^2 + 2\gamma^2)\Lambda}{(v_{\star}^2 \Lambda^2 + \gamma^2)^{3/2}}, \\
    \frac{d\ln \abs{\gamma}}{d\Lambda}&=-W_1 \frac{1}{4\pi} \frac{\Lambda}{(v_{\star}^2 \Lambda^2 + \gamma^2)^{1/2}}. \label{eq:gamma_flow}
\end{align}
The results show that both $v_\star$ and $\gamma$ increase as $\Lambda$ decreases under RG.
The RG equations for the interaction couplings (to leading order) are:
\begin{align}\label{eq:dV_lead}
    \frac{dV}{d\Lambda} &=  \mathcal{O}\qty(\frac{\gamma^4}{v_{\star}^4 \Lambda^4}),\\
     \frac{d U}{d\Lambda} &=  \frac{\qty[(W_1 - W_3)^2 + \frac{1}{4} J(W_1 - W_3)]}{\pi v_{\star}} + \mathcal{O}\qty(\frac{\gamma^2}{v_{\star}^2 \Lambda^2}), \\
     \frac{d W_1}{d\Lambda} &= -\frac{1}{8\pi v_{\star}} V(W_1 - W_3) + \mathcal{O}\qty(\frac{\gamma^2}{v_{\star}^2 \Lambda^2}), \\
     \frac{d W_3}{d\Lambda} &= \frac{1}{8 \pi v_{\star}} V(W_1 - W_3) + \mathcal{O}\qty(\frac{\gamma^2}{v_{\star}^2 \Lambda^2}), \\
     \frac{d J}{d\Lambda} &= -\frac{1}{8 \pi v_{\star}} VJ + \mathcal{O}\qty(\frac{\gamma^2}{v_{\star}^2 \Lambda^2}), \\
     \frac{d J_{+}}{d\Lambda} &= \frac{1}{8 \pi v_{\star}} VJ_{+}  + \mathcal{O}\qty(\frac{\gamma^2}{v_{\star}^2 \Lambda^2}).\label{eq:dJp_lead}
\end{align}
The detailed derivation of the RG Eqs.~(\ref{eq:v_star_flow}-\ref{eq:dJp_lead}) are given in Sec.~\ref{sec:derivation_flow}, and the full RG equations including all subleading one-loop contributions are given in Appendix~\ref{sec:RG_beyond_leading}. 
Here, the interaction strengths $V,U,W,J,J_{+}$ are in units of meV$\cdot \Omega_0$, where $\Omega_0 = a_M^2\sqrt{3}/2$ represents the area of a moiré unit cell, $a_M=13.4$ nm is the moiré lattice constant at the magic angle $\theta=1.05\degree$.
For convenience, we set $\Omega_0\to1$ in the RG equations and the corresponding RG flow plots in Figs.~\ref{fig:gamma_vstar_RG} and \ref{fig:interaction_flow}.
$\Lambda$ labels the momentum of remote band electrons, above which we integrate out in the partition function.
The (ultraviolet) initial conditions at $\Lambda_c=4\pi/3a_M$ for the single-particle parameters and interaction strengths are chosen to match the THF model corresponding to $w_0/w_1=0.8$ and $\theta=1.05\degree$ with projected Coulomb interaction screened by a dual gate separated by 10 nm hBN from the MATBG and a lattice dielectric constant $\kappa =6$~\cite{Song:2022}.
The initial values at $\Lambda_c$ are $v_{\star}=4.303$ eV$\cdot$\AA, $\gamma=24.75$ meV, $V=48.31$ meV, $U=57.95$ meV, $W_1=44.03$ meV, $W_3=50.20$ meV, $J=J_{+}=16.38$ meV \cite{Song:2022}.
For systems with different $\theta$, $w_0/w_1$, and $d$, the initial interaction values at the cutoff $\Lambda_c$ should be adjusted accordingly, as discussed in Ref.~\cite{Calugaru:2023}.

Numerical integration of these equations, using parameters fitted from the BM model with $w_0/w_1 = 0.8$ and $\theta = 1.05^\circ$, shows that $v_{\star}$ and $\gamma$ increase by approximately 30\% over the RG flow window (see Fig.~\ref{fig:gamma_vstar_RG}). 
Importantly, this RG flow in the THF model is consistent with the trend toward the chiral limit in the BM model~\cite{Vafek:2020}, where the AA/AB tunneling ratio $w_0/w_1$ decreases. 
Fitting (non-interacting) THF model parameters from BM calculations with varying $w_0/w_1$ confirms that both $v_{\star}$ and $\gamma$ increase as $w_0/w_1$ is lowered (see, e.g., Table S4 in Ref.~\cite{Song:2022} and Table S38 in Ref.~\cite{Calugaru:2023}), consistent with the upward flow observed in our RG treatment. 
A similar qualitative increase in $\gamma$ has also been reported in a self-consistent Hartree–Fock study using a tight-binding model~\cite{Miguel:2025}.
We emphasize that this tendency to approach the chiral limit of the BM model is not an artifact of setting $M$ and $v_{\smallstar}$ to zero. 

On the other hand, for the interaction strengths, we observe the following qualitative trends (see Fig.~\ref{fig:interaction_flow}).
$U$ remains nearly constant throughout the RG flow, validating the use of its UV values for low-energy modeling~\cite{Song:2022,Calugaru:2023,Rai:2024,Chou:2023a,Hu:2023a,Hu:2023b,Zhou:2024,Chou:2023b,Hu:2025,Zhao:2025}.
$V$ and $W$ also experience only weak renormalization (on the order of 10\%).
On the other hand, $J$ increases while $J_{+}$ decreases, leading to a roughly doubling of the ratio $J/J_{+}$ in the flow window. 
This reflects the enhanced importance of exchange interactions over double-hybridization-assisted processes in the low-energy effective Hamiltonian.

Since $\gamma$ increases while $U$ remains nealy unchanged under RG, this flow leads to a decreasing ratio $U/\gamma$ as the cutoff $\Lambda$ is lowered.
Theoretically, in the limit of $v_{\star} \Lambda_c \gg U,\gamma$, this RG flow effectively drives the system towards the projected-limit/Mott-semimetal regime where $U\ll \gamma$, even given $U > \gamma$ at the UV scale.
In practice, the ratios $v_{\star} \Lambda_c/U \sim 3$ and $v_{\star} \Lambda_c/\gamma \sim 5$ are finite, so the resulting energy hierarchy depends on the initial conditions.
However, the trend that $U/\gamma$ decreases under RG still holds.
\begin{figure}
    \centering
    \includegraphics[width=0.8\linewidth]{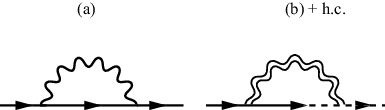}
    \caption{Leading non-zero self-energy correction.}
    \label{fig:self_energy_leading}
\end{figure}
\begin{figure}
    \centering
    \includegraphics[width=0.8\linewidth]{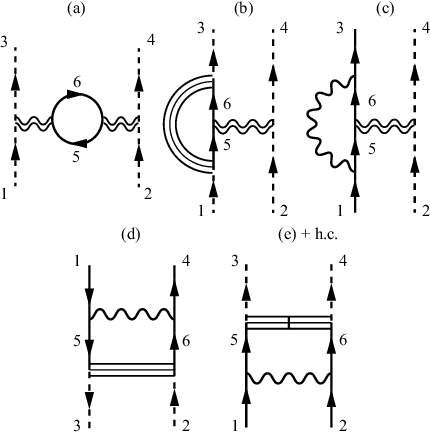}
    \caption{Leading non-zero one-loop diagrams for interaction correction.}
    \label{fig:interaction_leading}
\end{figure}

\subsection{Derivation of Flow Equations}
\label{sec:derivation_flow}
\begin{figure}
    \centering
    \includegraphics[width=0.8\linewidth]{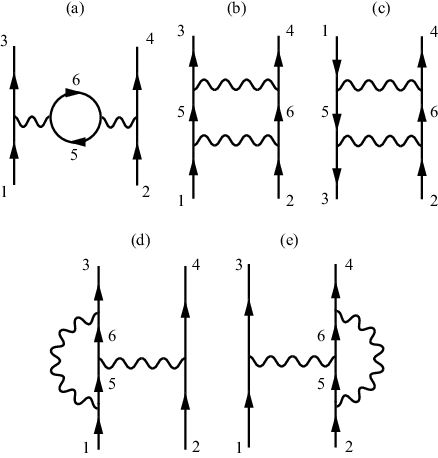}
    \caption{$c$-fermion (or $V$) only one-loop diagram. }
    \label{fig:V_c_only}
\end{figure}

Here, we compute the one-loop diagrams shown in Fig.~\ref{fig:interaction_leading}-\ref{fig:self_energy} and derive our results of RG flows given by Eqs.~(\ref{eq:v_star_flow}-\ref{eq:dJp_lead}).
Throughout the calculation, we take the external momenta to be small compared to internal momenta in the shell, a standard approximation in momentum-shell RG, since we are interested in the low-energy physics.
For example, when we compute the exchange self-energy, we are going to evaluate integral such as
\begin{align}\label{eq:exchange_integral}
    \int' \frac{d^2\vb{q}}{(2\pi)^2} \int_{-\infty}^{\infty}\frac{d\nu}{2\pi} G^{\alpha \eta}_1 (i\omega + i\nu, \vb{k}+\vb{q}),
\end{align}
where $\{\omega,\vb{k}\}$ are external and $\{\nu,\vb{q}\}$ are internal, and $\int'$ indicates the integral range is over the momentum shell $\Lambda-d\Lambda < \abs{\vb{q}}<\Lambda$.
On the other hand, the bubble self-energy diagrams involve integral $\int'_q G_1^{\alpha\eta}(q)$, which vanishes except the matrix element between $f_{\alpha}$ and $c_{\alpha}$:
\begin{align}\label{eq:int_G}
    \int'_q G_1^{\alpha \eta}(q) 
    = -\frac{\Lambda d\Lambda}{4\pi \sqrt{v_{\star}^2 \Lambda^2 + \gamma^2}} 
    \begin{bmatrix}
        0 & \gamma & 0 \\
        \gamma & 0 & 0 \\
        0 & 0 & 0
    \end{bmatrix}.
\end{align}
However, for the interactions Eqs.~(\ref{eq:S_U}-\ref{eq:S_J+}) [cf. Fig.~\ref{fig:Coulomb_interactions}], there is no vertex connecting $f_{\alpha}$ and $c_{\alpha}$, so there is no bubble self-energy correction. Note that the interactions $J$ and $J_+$ [Eqs.~(\ref{eq:S_J}-\ref{eq:S_J+})] involve $f_{\alpha}$ and $c_{\alpha+2}$ (not $c_{\alpha}$). 
To compute the exchange integral, Eq.~\eqref{eq:exchange_integral}, one can expand the Green function $G_1(k+q)$ to the leading order in $k$ assuming $k\ll q$: 
\begin{align}
    G^{\alpha \eta}_1 (z+x, \vb{k}+\vb{q}) \approx G^{\alpha \eta}_1 (x, \vb{q}) + \delta G^{\alpha\eta}(x, \vb{q}; z, \vb{k}).
\end{align}
The full expressions of $\delta G$ and the corresponding integrals are shown in Appendix~\ref{sec:Green_function_identities}.
This leads to 
\begin{align}
    &\int'_q \delta G^{\alpha \eta}(q,k) \approx 
    -\frac{\Lambda d\Lambda}{8\pi} \frac{(v_{\star}^2\Lambda^2+2\gamma^2) }{( v_{\star}^2 \Lambda^2+\gamma^2)^{3/2}} \times \nonumber \\
    & \times\begin{bmatrix}
        0 & 0 & 0 \\
        0 & 0 &  v_{\star} \abs{\vb{k}} \eta e^{i \eta s_{\alpha} \theta_{\vb{k}}} \\
        0 &v_{\star} \abs{\vb{k}} \eta e^{-i \eta s_{\alpha} \theta_{\vb{k}}} & 0
    \end{bmatrix}. \label{eq:int_delta_G}
\end{align}
As a result, only the exchange diagrams in Fig.~\ref{fig:self_energy_leading} have non-zero contribution.
The exchange diagram in Fig.~\ref{fig:self_energy_leading} (a) reads
\begin{align}
    \delta S^{(1)}_{(\mathrm{a})}&=- \int_k {\bar c}_{k, a, \eta, s} {c}_{k, a', \eta, s} \int_{q}'  V G^{(c)\eta}_{a,a'}(k+q), \\
    &\approx - \int_k {\bar c}_{k, a, \eta, s} {c}_{k, a',  \eta, s} \int_{q}' V \delta G^{(c)\eta}_{a,a'} (q,k),
\end{align}
which contributes to the renormalization of $c$-fermion velocity $v_{\star}$ [cf. Eqs.~(\ref{eq:v_star_flow}, \ref{eq:int_delta_G})]:
\begin{align}
    \delta v_{\star} &= v_{\star}(\Lambda-d\Lambda) - v_{\star}(\Lambda),\nonumber \\
    &= \int' \frac{d^2\vb{q}}{(2\pi)^2} \int_{-\infty}^{\infty} \frac{d\nu}{2\pi} \frac{(\nu^2 +\gamma^2) v_{\star}}{(\nu^2 + v_{\star}^2 \abs{\vb{q}}^2 + \gamma^2)}, \nonumber \\
    &= v_{\star} \frac{V}{8\pi} \frac{(v_{\star}^2 \Lambda^2 + 2\gamma^2)\Lambda d \Lambda}{(v_{\star}^2 \Lambda^2 + \gamma^2)^{3/2}}.
\end{align}
Similarly, the exchange diagram in Fig.~\ref{fig:self_energy_leading} (b) reads
\begin{align}
    \delta S^{(1)}_{(\mathrm{b})}&=- \int_k {\bar f}_{k, \alpha, \eta, s} {c}_{k, a, \eta, s} \int_{q}'  W_a G^{(fc)\eta}_{\alpha,a}(k+q), \\
    &\approx - \int_k {\bar f}_{k,\alpha \eta s} {c}_{k, a, \eta, s} \int_{q}'  W_a G^{(fc)\eta}_{\alpha,a}(q),
\end{align}
which contributes to the renormalization of $\gamma$ [cf. Eqs.~(\ref{eq:gamma_flow}, \ref{eq:int_G}-\ref{eq:int_delta_G})]:
\begin{align}
    \delta\gamma &= \gamma(\Lambda-d\Lambda)-\gamma(\Lambda)= -\int_{q}'W_a G^{(fc)\eta}_{\alpha,a}(q), \nonumber \\
    &= \int' \frac{d^2\vb{q}}{(2\pi)^2} \int_{-\infty}^{\infty} \frac{d\nu}{2\pi}\frac{\delta_{\alpha,a} W_1\gamma}{\nu^2 + v_{\star}^2 \abs{\vb{q}}^2 + \gamma^2}, \nonumber \\
    &= \gamma  \frac{W_1}{4\pi} \frac{\Lambda d\Lambda}{(v_{\star}^2 \Lambda^2 + \gamma^2)^{1/2}}.
\end{align}

Next, we compute the leading one-loop diagrams for interactions shown in Figs.~\ref{fig:interaction_leading} and \ref{fig:V_c_only} by treating $\gamma/(v_{\star} \Lambda)$ as a small parameter. 
By setting the external momentum to zero, we have the diagrams in Fig.~\ref{fig:interaction_leading}
\begin{align}
    \delta S^{(2)}_{\mathrm{(a)}} \approx &\frac{1}{2}\int_{\{k_i\}}\bar{f}_{k_3, \alpha, \eta, s} \bar{f}_{k_4,\alpha', \eta', s'} f_{k_2, \alpha', \eta', s'} f_{k_1, \alpha, \eta, s} \times\nonumber\\ 
    &\times \int_q' \sum_{\eta,s,a_1,a_2}G^{(c)\eta}_{a_1,a_2}(q) G^{(c)\eta}_{a_2,a_1}(q) W_{a_1} W_{a_2}, \\
    \delta S^{(2)}_{\mathrm{(b)}} \approx &-\frac{1}{2} \times 2\int_{\{k_i\}}\bar{f}_{k_3, \alpha, \eta, s} \bar{f}_{k_4,\alpha', \eta', s'} f_{k_2, \alpha', \eta', s'} f_{k_1, \alpha, \eta, s} \times\nonumber\\ 
    &\times \int_q' \sum_a G^{(c)\eta}_{a,\alpha'+2}(q) G^{(c)\eta}_{\alpha'+2,a}(q) W_a J, \\
    \delta S^{(2)}_{\mathrm{(c)}} \approx &-\frac{1}{2} \times 2 \int_{\{k_i\}}\bar{c}_{k_3, a, \eta, s} \bar{f}_{k_4,\alpha , \eta', s'} f_{k_2, \alpha, \eta', s'} c_{k_1, a, \eta, s} \times\nonumber\\
    &\times \int_q' \sum_{a_1} G^{(c)\eta}_{a_1,a}(q) G^{(c)\eta}_{a,a_1}(q) V W_{a_1}, \\
    \delta S^{(2)}_{\mathrm{(d)}} \approx &-\frac{1}{2} \times 2 \int_{\{k_i\}}\bar{f}_{k_3, \alpha, \eta, s} \bar{c}_{k_4, a', \eta', s'} f_{k_2, \alpha', \eta', s'} c_{k_1, a, \eta, s} \times\nonumber\\
    &\times \int_q' G^{(c)\eta}_{\alpha+2,a}(q) G^{(c)\eta'}_{a',\alpha'+2}(q) V J \frac{1}{2}(\eta \eta' + s_{\alpha} s_{\alpha'}), \\
    \delta S^{(2)}_{\mathrm{(e)}} \approx &-\frac{1}{2} \times 2 \int_{\{k_i\}} \bar{f}_{k_3, \alpha, \eta, s} \bar{f}_{k_4, \alpha', \eta', s'} c_{k_2, a', \eta', s'} c_{k_1, a, \eta, s} \times\nonumber\\
    &\times \int_q' G^{(c)\eta}_{\alpha+2,a}(q) G^{(c)\eta'}_{\alpha'+2,a'}(-q) V (-J_{+}) \frac{1}{2}(\eta \eta' - s_{\alpha} s_{\alpha'}),
\end{align}
which corresponds to the following interaction correction
\begin{align}
    \delta U^{\mathrm{(a)}} &= 8 \int' \frac{d^2\vb{q}}{(2\pi)^2}\int_{-\infty}^{\infty} \frac{d\nu}{2\pi} \times \nonumber \\
    &\times \frac{2v_{\star}^2\abs{\vb{q}}^2 W_1 W_3 - \nu^2 W_1^2 - \frac{\nu^2 v_{\star}^4\abs{\vb{q}}^4 W_3^2}{(v_{\star}^2\abs{\vb{q}}^2 + \gamma^2)^2}}{(\nu^2 + v_{\star}^2\abs{\vb{q}}^2 + \gamma^2)^2}, \nonumber \\
    &= - \frac{1}{\pi} \Lambda d\Lambda \frac{[v_{\star}^2 \Lambda^2 (W_1 - W_3) + W_1 \gamma^2]^2}{(v_{\star}^2 \Lambda^2 + \gamma^2)^{5/2}}, \\
    \delta U^{\mathrm{(b)}} &= \int' \frac{d^2\vb{q}}{(2\pi)^2}\int_{-\infty}^{\infty} \frac{d\nu}{2\pi} J \frac{v_{\star}^2\abs{\vb{q}}^2 \qty(W_1 - \frac{\nu^2 v_{\star}^2\abs{\vb{q}}^2 W_3}{(v_{\star}^2\abs{\vb{q}}^2 + \gamma^2)^2})}{(\nu^2 + v_{\star}^2\abs{\vb{q}}^2 + \gamma^2)^2}\nonumber \\
    &= - J \frac{1}{4\pi} \Lambda d\Lambda \frac{v_{\star}^2 \Lambda^2[v_{\star}^2 \Lambda^2 (W_1 - W_3) + W_1 \gamma^2]}{(v_{\star}^2 \Lambda^2 + \gamma^2)^{5/2}} ,
\end{align}
\begin{align}
    \delta W^{\mathrm{(c)}}_1 &= V\int' \frac{d^2\vb{q}}{(2\pi)^2}\int_{-\infty}^{\infty} \frac{d\nu}{2\pi} \frac{v_{\star}^2\abs{\vb{q}}^2 W_3 - \nu^2 W_1}{(\nu^2 + v_{\star}^2\abs{\vb{q}}^2 + \gamma^2)^2}, \nonumber \\
    &= V \frac{1}{8\pi} \Lambda d\Lambda \frac{v_{\star}^2 \Lambda^2 (W_1 - W_3) + W_1 \gamma^2}{(v_{\star}^2 \Lambda^2 + \gamma^2)^{3/2}}, \\
    \delta W^{\mathrm{(c)}}_3 &= V\int' \frac{d^2\vb{q}}{(2\pi)^2}\int_{-\infty}^{\infty} \frac{d\nu}{2\pi} \frac{v_{\star}^2\abs{\vb{q}}^2 \qty[W_1 - \frac{\nu^2 v_{\star}^2\abs{\vb{q}}^2 W_3}{(v_{\star}^2\abs{\vb{q}}^2 + \gamma^2)^2}]}{(\nu^2 + v_{\star}^2\abs{\vb{q}}^2 + \gamma^2)^2}, \nonumber \\
    &=-V \frac{1}{8\pi} \Lambda d\Lambda \frac{v_{\star}^2 \Lambda^2[v_{\star}^2 \Lambda^2 (W_1 - W_3) + W_1 \gamma^2]}{(v_{\star}^2 \Lambda^2 + \gamma^2)^{5/2}}.
\end{align}
As for the integral in $\delta S^{(2)}_{\mathrm{(d)}}$ and $\delta S^{(2)}_{\mathrm{(e)}}$, we need to discuss the indices of $a$ and $a'$.
If $a=\alpha+ 2$ and $a'=\alpha'+2$, then they contribute to the following interaction corrections
\begin{align}
    \delta J^{(d)} &= VJ \left[\int' \frac{d^2\vb{q}}{(2\pi)^2}\int_{-\infty}^{\infty} \frac{d\nu}{2\pi} \times \right.\nonumber \\
    &\left.\times\frac{\nu^2 v_{\star}^4 \abs{\vb{q}}^4}{(\nu^2 + v_{\star}^2\abs{\vb{q}}^2 + \gamma^2)^2(v_{\star}^2\abs{\vb{q}}^2 + \gamma^2)^2}\right], \nonumber\\
    &= VJ \frac{1}{8\pi} \Lambda d\Lambda \frac{v_{\star}^4 \Lambda^4}{(v_{\star}^2 \Lambda^2 + \gamma^2)^{5/2}}, \\
    \delta J_{+}^{(e)} &=-VJ_{+} \left[\int' \frac{d^2\vb{q}}{(2\pi)^2}\int_{-\infty}^{\infty} \frac{d\nu}{2\pi} \times \right. \nonumber \\
    &\left.\times\frac{\nu^2 v_{\star}^4 \abs{\vb{q}}^4}{(\nu^2 + v_{\star}^2\abs{\vb{q}}^2 + \gamma^2)^2(v_{\star}^2\abs{\vb{q}}^2 + \gamma^2)^2}\right], \nonumber\\
    &= - VJ_{+} \frac{1}{8\pi} \Lambda d\Lambda \frac{v_{\star}^4 \Lambda^4}{(v_{\star}^2 \Lambda^2 + \gamma^2)^{5/2}},
\end{align}
While if $a=\alpha$ and $a'=\alpha'$, then
they contribute to new interactions of the form 
\begin{align}\label{eq:S_Jstar}
    &S_{J^{\star}} = J^{\star}\frac{1}{2}(1 + \eta \eta' s_{\alpha} s_{\alpha'}) \times \nonumber \\
    &\times  \int_{\{k_i\}}{\bar f}_{k_3,\alpha,\eta,s} {\bar c}_{k_4,\alpha',\eta',s'} f_{k_2,\alpha',\eta',s'}  c_{k_1,\alpha,\eta,s}, \\
    &S_{J_{+}^{\star}} = -\frac{J_{+}^{\star}}{2}\frac{1}{2}(1 - \eta \eta' s_{\alpha} s_{\alpha'}) \times \nonumber \\
    &\times \int_{\{k_i\}} \qty({\bar f}_{k_3,\alpha,\eta,s} {\bar f}_{k_4,\alpha',\eta',s'} c_{k_2,\alpha',\eta',s'} c_{k_1,\alpha,\eta,s} + \mathrm{h.c.}),  \label{eq:S_Jp_star}
\end{align}
which can be seen from the integrals
\begin{align}
    &\int_{0}^{2\pi} \frac{d\theta_{\vb{q}}}{2\pi} G^{(c)\eta}_{\alpha+2,\alpha}(q) G^{(c)\eta'}_{\alpha',\alpha'+2}(q) \nonumber \\
    &= \frac{\eta \eta' v_{\star}^2\abs{\vb{q}}^2}{(\nu^2 + v_{\star}^2\abs{\vb{q}}^2 + \gamma^2)^2} \int_{0}^{2\pi} \frac{d\theta_{\vb{q}}}{2\pi} d\theta_{\vb{q}} e^{i(\eta s_{\alpha} - \eta' s_{\alpha'}) \theta_{\vb{q}}} \nonumber \\
    &= \frac{v_{\star}^2\abs{\vb{q}}^2}{(\nu^2 + v_{\star}^2\abs{\vb{q}}^2 + \gamma^2)^2} \frac{1}{2}(\eta \eta' + s_{\alpha} s_{\alpha'}),
\end{align}
and 
\begin{align}
    &\int_{0}^{2\pi} \frac{d\theta_{\vb{q}}}{2\pi} G^{(c)\eta}_{\alpha+2,\alpha}(q) G^{(c)\eta'}_{\alpha'+2,\alpha'}(-q) \nonumber \\
    &= \frac{-\eta \eta' v_{\star}^2\abs{\vb{q}}^2}{(\nu^2 + v_{\star}^2\abs{\vb{q}}^2 + \gamma^2)^2} \int_{0}^{2\pi} \frac{d\theta_{\vb{q}}}{2\pi} d\theta_{\vb{q}} e^{i(\eta s_{\alpha} + \eta' s_{\alpha'}) \theta_{\vb{q}}} \nonumber \\
    &= \frac{-v_{\star}^2\abs{\vb{q}}^2}{(\nu^2 + v_{\star}^2\abs{\vb{q}}^2 + \gamma^2)^2} \frac{1}{2}(\eta \eta' - s_{\alpha} s_{\alpha'}),
\end{align}
As a result, they contribute to the orbital-valley texture factor that appears in Eqs.~(\ref{eq:S_Jstar}-\ref{eq:S_Jp_star})
\begin{align}
    \qty[\frac{1}{2}(\eta \eta' + s_{\alpha} s_{\alpha'})]^2 = \frac{1}{2}(1 + \eta \eta' s_{\alpha} s_{\alpha'}), \\
    \qty[\frac{1}{2}(\eta \eta' - s_{\alpha} s_{\alpha'})]^2 = \frac{1}{2}(1 - \eta \eta' s_{\alpha} s_{\alpha'}).
\end{align}
The initial conditions for these new interactions are $J^{\star} = J_{+}^{\star} = 0$, and the corrections from $\delta S^{(2)}_{\mathrm{(d)}}$ and $\delta S^{(2)}_{\mathrm{(e)}}$ are given by
\begin{align}\label{eq:Jstar}
    \delta J^{\star} &= V J\frac{1}{8\pi} \Lambda d\Lambda  \frac{v_{\star}^2 \Lambda^2}{(v_{\star}^2 \Lambda^2 + \gamma^2)^{3/2}},\\
    \delta J_{+}^{\star} &= -V J_{+}\frac{1}{8\pi} \Lambda d\Lambda  \frac{v_{\star}^2 \Lambda^2}{(v_{\star}^2 \Lambda^2 + \gamma^2)^{3/2}}. \label{eq:J_plus_star}
\end{align}
In principle, one should account for all interaction terms generated during the RG flow and solve the full set of flow equations self-consistently [e.g., $J^{\star}$ also generates a correction to $\gamma$ through Eq.~\eqref{eq:int_G}]. 
However, in practice, the newly generated interactions are typically small in magnitude and can be neglected without significantly affecting the results. 
For example, one can approximately integrate Eqs.~(\ref{eq:Jstar}-\ref{eq:J_plus_star}) by holding $V$, $J$, and $J_{+}$ fixed at their initial values to estimate the renormalized couplings $J^{\star}$ and $J_{+}^{\star}$, which are found to be on the order of $\sim 1$~meV—an order of magnitude smaller than the bare values of $J$ and $J_{+}$.

On the other hand, the diagrams in Fig.~\ref{fig:V_c_only} involves only $c$-fermions and renormalize the $V$ term:
\begin{align}
    \delta S^{(2)}_{V\mathrm{(a)}} \approx &\frac{1}{2}\int_{\{k_i\}}\bar{c}_{k_3, a, \eta, s} \bar{c}_{k_4,a', \eta', s'} c_{k_2, a', \eta', s'} c_{k_1, a, \eta, s} \times\nonumber\\ 
    &\times \int_q' \sum_{\eta,s,a_1,a_2}G^{(c)\eta}_{a_1,a_2}(q) G^{(c)\eta}_{a_2,a_1}(q) V^2, \\
    \delta S^{(2)}_{V\mathrm{(b)}} \approx &-\frac{1}{2} \int_{\{k_i\}}\bar{c}_{k_3, b, \eta, s} \bar{c}_{k_4,b', \eta', s'} c_{k_2, a', \eta', s'} c_{k_1, a, \eta, s} \times\nonumber\\ 
    &\times \int_q' \sum_a G^{(c)\eta}_{b',b}(q) G^{(c)\eta}_{a',a}(-q) V^2, \\
    \delta S^{(2)}_{V\mathrm{(c)}} \approx &-\frac{1}{2} \int_{\{k_i\}}\bar{c}_{k_3, b, \eta, s} \bar{c}_{k_4,b', \eta', s'} c_{k_2, a', \eta', s'} c_{k_1, a, \eta, s} \times\nonumber\\ 
    &\times \int_q' \sum_a G^{(c)\eta}_{b',b}(q) G^{(c)\eta}_{a',a}(q) V^2, \\
    \delta S^{(2)}_{V\mathrm{(d)}} \approx &-\frac{1}{2} \int_{\{k_i\}}\bar{c}_{k_3, b, \eta, s} \bar{c}_{k_4,a', \eta', s'} c_{k_2, a', \eta', s'} c_{k_1, a, \eta, s} \times\nonumber\\ 
    &\times \int_q' \sum_a G^{(c)\eta}_{a_1,a}(q) G^{(c)\eta}_{b,a_1}(q) V^2, \\
    \delta S^{(2)}_{V\mathrm{(e)}} \approx &-\frac{1}{2} \int_{\{k_i\}}\bar{c}_{k_3, a, \eta, s} \bar{c}_{k_4,b', \eta', s'} c_{k_2, a', \eta', s'} c_{k_1, a, \eta, s} \times\nonumber\\ 
    &\times \int_q' \sum_a G^{(c)\eta}_{a_1,a'}(q) G^{(c)\eta}_{b',a_1}(q) V^2.
\end{align}
We notice that $S^{(2)}_{V\mathrm{(b)}} =-S^{(2)}_{V\mathrm{(c)}}$ since the Green function is an odd function of the momentum $G^{(c)\eta}_{a',a}(-q) = -G^{(c)\eta}_{a',a}(q)$, so they cancel each other. 
The remaining diagrams corresponds to the following interaction correction
\begin{align}
    \delta V^{(a)} &= 8V^2 \int' \frac{d^2\vb{q}}{(2\pi)^2}\int_{-\infty}^{\infty} \frac{d\nu}{2\pi} \frac{-\nu^2 + 2v_{\star}^2\abs{\vb{q}}^2 + \frac{-\nu^2v_{\star}^4\abs{\vb{q}}^4}{(v_{\star}^2\abs{\vb{q}}^2+\gamma^2)^2}}{(\nu^2 + v_{\star}^2\abs{\vb{q}}^2+\gamma^2)^2}, \nonumber \\
    &= - \frac{V^2}{\pi}  \frac{\gamma^4 \Lambda d\Lambda}{(v_{\star}^2 \Lambda^2 + \gamma^2)^{5/2}}, 
\end{align}
\begin{align}
    &\delta V^{(d)}_{ab} = \delta V^{(e)}_{a'b'} = V^2 \int' \frac{d^2\vb{q}}{(2\pi)^2}\int_{-\infty}^{\infty} \frac{d\nu}{2\pi} \times \nonumber \\
    &\times \frac{1}{(\nu^2 + v_{\star}^2\abs{\vb{q}}^2+\gamma^2)^2}
    \begin{bmatrix}
        -\nu^2 + v_{\star}^2\abs{\vb{q}}^2 & \sim \nu e^{i\theta_{\vb{q}}} v_{\star}\abs{\vb{q}} \\
        \sim \nu e^{-i\theta_{\vb{q}}} v_{\star}\abs{\vb{q}} & v_{\star}^2\abs{\vb{q}}^2 + \frac{-\nu^2 v_{\star}^4\abs{\vb{q}}^4}{v_{\star}^2\abs{\vb{q}}^2 + \gamma^2}
    \end{bmatrix}, \\
    &\to \delta V^{(d,e)} = \frac{1}{2} \mathrm{Tr} [\delta V^{(d,e)}_{ab}] = \frac{V^2}{8 \pi} \frac{\gamma^4 \Lambda d\Lambda}{(v_{\star}^2 \Lambda^2 + \gamma^2)^{5/2}}.
\end{align}
As a result, the total contribution from the $c$-fermion-only diagrams $\propto (\gamma/v_{\star}\Lambda)^4$ vanishes to the leading order compared to other diagrams in Fig.~\ref{fig:interaction_leading}, which agrees with previous results in graphene systems~\cite{Foster_Aleiner:2008,Vafek_Yang:2010}. 
The resulting analytical expressions for the leading order RG flows, given in Eqs.~(\ref{eq:v_star_flow}-\ref{eq:dJp_lead}), describe how the interplay between hybridization and Coulomb interactions shapes the low-energy effective theory of MATBG in the THF framework.


\section{Conclusion}
\label{sec:conclusion}

In this work, we have developed a perturbative RG framework based on the THF model to investigate the interplay of correlations and topology in MATBG.  By focusing on the experimentally relevant intermediate regime where the flat-band Coulomb interaction scale $U$ is comparable to the hybridization strength $\gamma$, our work provides useful insight for understanding the evolution of low-energy parameters in a topologically nontrivial, strongly correlated moiré system.

Our RG analysis reveals that while the single-particle parameters $v_{\star}$ and $\gamma$ flow upward with decreasing energy scale, the interaction strength $U$ remains approximately constant, resulting in a decreasing ratio $U/\gamma$. This trend suggests that, even when starting from a regime with $U > \gamma$, the system may naturally flow toward a projected-limit/Mott-semimetal regime where $U < \gamma$.

Our RG results mirror the THF band-structure fits using the BM model: as the ratio $w_0/w_1$ decreases, both $\gamma$ and $v_{\star}$ increases (in line with Table S4 in Ref.~\cite{Song:2022} and Table S38 in Ref.~\cite{Calugaru:2023}). For example, comparing the hybridization gap before ($\gamma=25$ meV) and after ($\gamma=33$ meV) renormalization shows that $w_0/w_1$ shifts only from 0.8 to $\approx 0.7$. While this flow nudges the system toward the chiral regime, it falls well short of the exact chiral limit $w_0/w_1=0$. The finite RG window below $v_{\star} \Lambda_c \approx 130$ meV limits how far the parameters can run. We stress that these numbers alone do not let us declare whether the chiral or the THF framework is the better starting point. Instead, our calculation highlights a concrete link between two seemingly disparate models and may offer clues for reconciling them.

Our results show that $U$ remains nearly constant throughout the RG flow, which may explain the consistency between its UV value and the value $\approx 50$ meV extracted from the STM experiments~\cite{Xie:2019,Jiang:2019,Choi:2019,Wong:2020,Choi:2021,Nuckolls:2023}.

Compared to previous RG approaches based on the BM model~\cite{Vafek:2020}, our THF-based formalism offers analytical transparency and computational efficiency. It allows a direct decomposition of interaction channels into physically interpretable forms and quantitatively tracks the evolution of these couplings. Notably, we find that the exchange interaction $J$ becomes more prominent relative to the double-hybridization term $J_{+}$ as the system flows toward lower energies. 

The interaction renormalization is minimal for the intermediate energy scales $|E|>\gamma$. The minimal effect stems from the Dirac-like density of states as discussed in several papers in graphene~\cite{Gonzalez:1994,Vafek:2007,Sheehy:2007,Barnes:2014}. However, significant interaction renormalization can happen at lower energies. For example, an earlier study based on the single impurity model predicts that the $J$ term becomes irrelevant away from the integer fillings for $|E|\gg\gamma$ \cite{Chou:2023b}. The Kondo interaction, which arises after projecting out local $f$-fermion charge fluctuation, also tends to govern the low-energy properties away from the integer fillings \cite{Zhou:2024,Chou:2023b}. Our work focuses on the complementary results at the intermediate energy scales, serving as a starting point and defining the bare parameters for the low-energy effective theory at scales $|E| < U$ and providing analytical understanding to the complicated interacting MATBG systems.

In our analysis, we neglect the momentum dependence of the interactions by assuming a screened short-range Coulomb potential. This approximation is justified when the characteristic length scale of interest is much larger than the screening length, which in our case is set by the distance to the metallic gate, $d$. Equivalently, the relevant momenta $\abs{\vb{k}}$ must satisfy $\abs{\vb{k}} \ll 1/d$. Since $d \sim 10$~nm is comparable to the moiré lattice constant $a_M \simeq 13$~nm, this condition holds for $\abs{\vb{k}} \ll 1/d \sim \Lambda_c$ within the first mBZ, justifying the neglect of umklapp processes.  
In contrast, when $d \gg a_M$, screening becomes ineffective, and the Coulomb interaction must be treated as long-range, scaling as $1/\abs{\vb{k}}$. In this regime, both the momentum dependence of the interaction and umklapp processes must be incorporated into the RG analysis. Notably, such a treatment can generate additional interactions, such as RKKY-like couplings between the $f$-fermions \cite{Hu:2023a,Hu:2025}. 
Other than gate screening, there is also screening from the electron. Electron screening may be implemented using static RPA~\cite{Foster_Aleiner:2008}, while the RG treatment of dynamically screened Coulomb interactions would require next‑to‑leading‑order diagrams in the RPA~\cite{Hofmann:2014}, which is beyond the scope of this paper.

While a fully self-consistent renormalization group calculation would track all interaction terms generated during the flow, in practice, these newly generated couplings--such as $J^{\star}$ and $J_+^{\star}$ defined in Eqs.~(\ref{eq:S_Jstar}-\ref{eq:S_Jp_star})--are found to be small in magnitude $\sim 1$ meV comparable to various other exchange terms that we ignore in the THF model~\cite{Song:2022}, and can be neglected without significantly altering the results. 
In addition, it is likely that the higher-loop effects should be negligible since the flow is minimal for interactions. 

Our RG framework is broadly applicable to other moir\'e systems and flat-band materials that can be effectively described by the THF model. Notable examples include magic-angle mirror-symmetric twisted trilayer graphene \cite{Yu:2023,Kim:2022,Banerjee:2025,Park:2025}, which differs from MATBG by featuring an additional Dirac band in the non-interacting spectrum; the twisted checkerboard model \cite{Ming-Rui_Li:2022,Sarkar:2023,Jia-Zheng_Ma:2024}; and the Lieb lattice~\cite{Lieb:1989,THF_principle_herzogarbeitman:2024}, all of which exhibit similar flat-band structures and topological features. More generally, our formalism is adaptable to stoichiometric systems and engineered lattices that host hybridized band structures resembling those of atomic $f$-electrons coupled to topological semimetals~\cite{THF_principle_herzogarbeitman:2024}, making it a versatile tool for exploring interaction-driven phenomena across a wide class of quantum materials.
Future directions include incorporating effects of finite doping away from CNP, strain, disorder, and comparing predictions with spectroscopic and transport experiments. These efforts will be important to establishing a unified low-energy description of correlated moiré flat bands and their emergent quantum phases.

\begin{acknowledgments}
We thank Matt Foster, Jay Sau, Fengcheng Wu, and Jiabin Yu for useful discussions. This work is supported by the Laboratory for Physical Sciences.
\end{acknowledgments}

\medskip
\onecolumngrid
\appendix
\section{Integrals and identities related to Green functions}
\label{sec:Green_function_identities}
In this appendix, we list some useful integrals and identities related to Green functions in the diagram calculation.
\subsection{Frequency integral}
If $a>0$ and $n>1/2$, then we have the integral
\begin{align}
    \int_{-\infty}^{\infty} \frac{d\omega}{(\omega^2 + a^2)^n} = \sqrt{\pi}\frac{a^{1-2n}  \Gamma(n-1/2)}{\Gamma(n)},
\end{align}
where $\Gamma(n)$ is the Gamma function.
This can be used to compute the integral appeared in many diagrams involving a product of two Green functions, for example,
\begin{align}
    \int_{-\infty}^{\infty} d\omega\frac{\gamma^2}{(\omega^2 + v_{\star}^2 k^2 + \gamma^2)^2} = \frac{\pi}{2} \frac{\gamma^2}{(v_{\star}^2 k^2 + \gamma^2)^{3/2}}.
\end{align}

\subsection{Trace in polarization bubble}
In the polarization bubble diagram such as Fig.~\ref{fig:interaction_leading} (a) and \ref{fig:V_c_only} (a), we encounter the trace of the Green functions
\begin{align}
    \sum_{a_1,a_2}G^{(c)\eta}_{a_1,a_2}(z,\vb{k}) G^{(c)\eta}_{a_2,a_1}(z,\vb{k}) W_{a_1} W_{a_2} &= \mathrm{Tr} 
    \left(
\begin{array}{cc}
 \frac{W_1 \left(v_{\star}^2 \abs{\vb{k}}^2 W_3+W_1 z^2\right)}{\left(\gamma ^2+v_{\star}^2 \abs{\vb{k}}^2-z^2\right)^2} & \frac{\eta  k v W_3 z e^{i s_\alpha  \eta  \theta } \left(v_{\star}^2 \abs{\vb{k}}^2 (W_1+W_3)+\gamma ^2 W_1\right)}{\left(\gamma ^2+v_{\star}^2 \abs{\vb{k}}^2\right) \left(\gamma ^2+v_{\star}^2 \abs{\vb{k}}^2-z^2\right)^2} \\[2mm]
 \frac{\eta  k v W_1 z e^{-i s_\alpha  \eta  \theta } \left(v_{\star}^2 \abs{\vb{k}}^2 (W_1+W_3)+\gamma ^2 W_1\right)}{\left(\gamma ^2+v_{\star}^2 \abs{\vb{k}}^2\right) \left(\gamma ^2+v_{\star}^2 \abs{\vb{k}}^2-z^2\right)^2} & \frac{v_{\star}^2 \abs{\vb{k}}^2 W_3 \left(\frac{v_{\star}^2 \abs{\vb{k}}^2 W_3 z^2}{\left(\gamma ^2+v_{\star}^2 \abs{\vb{k}}^2\right)^2}+W_1\right)}{\left(\gamma ^2+v_{\star}^2 \abs{\vb{k}}^2-z^2\right)^2} \\
\end{array}
\right) \nonumber \\
&= \frac{2 v_{\star}^2 \abs{\vb{k}}^2 W_1 W_3+\frac{v_{\star}^4 \abs{\vb{k}}^4 W_3^2 z^2}{\left(\gamma ^2+v_{\star}^2 \abs{\vb{k}}^2\right)^2}+W_1^2 z^2}{\left(\gamma ^2+v_{\star}^2 \abs{\vb{k}}^2-z^2\right)^2},
\end{align}
\begin{align}
    \sum_{a_1,a_2}G^{(c)\eta}_{a_1,a_2}(z,\vb{k}) G^{(c)\eta}_{a_2,a_1}(z,\vb{k})  &= \frac{1}{\left(\gamma ^2+v_{\star}^2 \abs{\vb{k}}^2-z^2\right)^2}\mathrm{Tr}\left(
\begin{array}{cc}
 v_{\star}^2 \abs{\vb{k}}^2+z^2 & \frac{\eta  k v z e^{i s_\alpha  \eta  \theta } \left(\gamma ^2+2 v_{\star}^2 \abs{\vb{k}}^2\right)}{\gamma ^2+v_{\star}^2 \abs{\vb{k}}^2} \\[1mm]
 \frac{\eta  k v z e^{-i s_\alpha  \eta  \theta } \left(\gamma ^2+2 v_{\star}^2 \abs{\vb{k}}^2\right)}{\gamma ^2+v_{\star}^2 \abs{\vb{k}}^2} & v_{\star}^2 \abs{\vb{k}}^2+\frac{v_{\star}^4 \abs{\vb{k}}^4 z^2}{\left(\gamma ^2+v_{\star}^2 \abs{\vb{k}}^2\right)^2} \\
\end{array}
\right) \nonumber \\
&= \frac{2 v_{\star}^2 \abs{\vb{k}}^2+\frac{v_{\star}^4 \abs{\vb{k}}^4 z^2}{\left(\gamma ^2+v_{\star}^2 \abs{\vb{k}}^2\right)^2}+z^2}{\left(\gamma ^2+v_{\star}^2 \abs{\vb{k}}^2-z^2\right)^2}.
\end{align}

\subsection{Integrals in self-energy diagrams}
The bubble self-energy diagram involves integral $\int'_q G_1(q)$. Since $G_1(q)$ is an odd function in $q$ except for the matrix element $G_{\alpha,\alpha}^{(cf)} (q)$ [cf. Eq.~\eqref{eq:G_1}], we have the following result
\begin{align}
    \int'_q G_1^{\alpha \eta}(q) &= \int_{\Lambda-d\Lambda}^{\Lambda} \frac{\abs{\vb{q}}d\abs{\vb{q}}}{2\pi} \int_{-\infty}^{\infty} \frac{d \nu}{2\pi}\int_0^{2\pi} \frac{d\theta_{\vb{q}}}{2\pi} G_1^{\alpha \eta}(q) \nonumber \\
    &= -\frac{\Lambda d\Lambda}{4\pi \sqrt{v_{\star}^2 \Lambda^2 + \gamma^2}} 
    \begin{bmatrix}
        0 & \gamma & 0 \\
        \gamma & 0 & 0 \\
        0 & 0 & 0
    \end{bmatrix}.
\end{align}
On the other hand, the exchange self energy diagram involves integral $\int_q' G_1(q+k)$ where $q$ is the internal momentum and $k$ is the external momentum.
To compute this integral, we expand the Green function assuming $k\ll q$:
\begin{align}
    G^{\alpha \eta}_1 (z+x, \vb{k}+\vb{q}) \approx G^{\alpha \eta}_1 (x, \vb{q}) + 
    \begin{bmatrix}
        \delta G^{(f)} & \delta G^{(fc)\eta} \\
        \delta G^{(cf)\eta} & \delta G^{(c)\eta}
    \end{bmatrix},
\end{align}
where
\begin{align}
    \delta G^{(f)} &= \frac{-z\gamma^2(x^2+v_{\star}^2 \abs{\vb{q}}^2+\gamma^2)}{(v_{\star}^2 \abs{\vb{q}}^2+\gamma^2)(-x^2 + v_{\star}^2 \abs{\vb{q}}^2+\gamma^2)^2}  + \nonumber \\
    &+ \frac{2 x \gamma ^2 v_{\star}^2 \vb{k} \cdot \vb{q} \left(2 \gamma ^2+2 v_{\star}^2 \abs{\vb{q}}^2 - x^2\right)}{\left(\gamma ^2+v_{\star}^2 \abs{\vb{q}}^2\right)^2 \left(-x^2+v_{\star}^2 \abs{\vb{q}}^2+\gamma ^2\right)^2},
\end{align}
\begin{align}
    &\delta G^{(c)\eta}=  
    \begin{bmatrix}
        \frac{-z (x^2+v_{\star}^2 \abs{\vb{q}}^2+\gamma^2)}{(-x^2 + v_{\star}^2 \abs{\vb{q}}^2+\gamma^2)^2} & \frac{-2xzv_{\star} \abs{\vb{q}} \eta e^{is_{\alpha} \eta \theta_{\vb{q}}}}{(-x^2 + v_{\star}^2 \abs{\vb{q}}^2+\gamma^2)^2} \\[1mm]
        \frac{-2xzv_{\star} \abs{\vb{q}} \eta e^{-is_{\alpha} \eta \theta_{\vb{q}}}}{(-x^2 + v_{\star}^2 \abs{\vb{q}}^2+\gamma^2)^2} & \frac{-zv_{\star}^2 \abs{\vb{q}}^2(x^2+v_{\star}^2 \abs{\vb{q}}^2+\gamma^2)}{(v_{\star}^2 \abs{\vb{q}}^2+\gamma^2) (-x^2 + v_{\star}^2 \abs{\vb{q}}^2+\gamma^2)^2}
    \end{bmatrix}+ \nonumber \\
    &+
    \begin{bmatrix}
        \frac{2 x v_{\star}^2 \vb{k} \cdot \vb{q}}{\left(-x^2+v_{\star}^2 \abs{\vb{q}}^2+\gamma^2\right)^2} & \frac{2 v_{\star}^3 \vb{k} \cdot \vb{q} (\eta q_x + i s_\alpha q_y)}{\left(-x^2+v_{\star}^2 \abs{\vb{q}}^2+\gamma^2\right)^2} - \frac{v_{\star} \abs{\vb{k}} \eta e^{i \eta s_{\alpha} \theta_{\vb{k}}}}{-x^2+v_{\star}^2 \abs{\vb{q}}^2+\gamma ^2} \\[1mm]
        \frac{2 v_{\star}^3 \vb{k} \cdot \vb{q} (\eta q_x - i s_\alpha q_y)}{\left(-x^2+v_{\star}^2 \abs{\vb{q}}^2+\gamma^2\right)^2} - \frac{v_{\star} \abs{\vb{k}} \eta e^{-i \eta s_{\alpha} \theta_{\vb{k}}}}{-x^2+v_{\star}^2 \abs{\vb{q}}^2+\gamma ^2} & \frac{2 x v^2 \vb{k} \cdot \vb{q} \left( x^2 \gamma^2+v_{\star}^4 \abs{\vb{q}}^4-\gamma ^4\right)}{\left(v_{\star}^2 \abs{\vb{q}}^2+ \gamma^2\right)^2 \left(-x^2+v_{\star}^2 \abs{\vb{q}}^2+\gamma ^2\right)^2}
    \end{bmatrix}
\end{align}
\begin{align}
    \delta G^{(fc)\eta} &= \delta G^{(cf)\eta \dagger} = 
    \begin{bmatrix}
        \frac{-2 xz\gamma  }{\left(-x^2+v_{\star}^2 \abs{\vb{q}}^2+\gamma ^2\right)^2}, & \frac{-z\gamma   v_{\star} \abs{\vb{q}} \eta e^{i s_\alpha  \eta  \theta_{\vb{q}} } \left(x^2+v_{\star}^2 \abs{\vb{q}}^2+\gamma ^2\right)}{\left(v_{\star}^2 \abs{\vb{q}}^2 + \gamma ^2\right) \left(-x^2+v_{\star}^2 \abs{\vb{q}}^2+\gamma ^2\right)^2}
    \end{bmatrix} +  \nonumber \\
    &+ 
    \begin{bmatrix}
        \frac{2 \gamma v_{\star}^2 \vb{k} \cdot \vb{q} }{\left(-x^2+v_{\star}^2 \abs{\vb{q}}^2+\gamma ^2\right)^2}, & x\gamma\eta \frac{ 2 v_{\star}^3 (\eta  q_x+i s_\alpha  q_y) \vb{k}\cdot\vb{q} \left(-x^2+2 v_{\star}^2 \abs{\vb{q}}^2+2 \gamma ^2\right)- v_{\star} (\eta  k_x+i s_\alpha  k_y) \left(\gamma ^2+v_{\star}^2 \abs{\vb{q}}^2\right) \left(-x^2+v_{\star}^2 \abs{\vb{q}}^2+\gamma^2\right)}{\left(v_{\star}^2 \abs{\vb{q}}^2 + \gamma^2\right)^2 \left(-x^2+v_{\star}^2 \abs{\vb{q}+\gamma ^2}^2\right)^2}
    \end{bmatrix}
\end{align}
Here, $z=i\omega$ and $x=i \nu$ represent the external and internal Matsubara frequency, respectively. To calculate the integral $\int'_{q}\delta G$, one can first compute the angle integral of $\theta_{\vb{q}}$, such that 
\begin{align}
    \int_0^{2\pi} \frac{d\theta_{\vb{q}}}{2\pi}\delta G^{(f)} &= \frac{-z\gamma^2(x^2+v_{\star}^2 \abs{\vb{q}}^2+\gamma^2)}{(v_{\star}^2 \abs{\vb{q}}^2+\gamma^2)(-x^2 + v_{\star}^2 \abs{\vb{q}}^2+\gamma^2)^2}, \\
    \int_0^{2\pi} \frac{d\theta_{\vb{q}}}{2\pi}\delta G^{(c)\eta} &= 
    \begin{bmatrix}
        \frac{-z (x^2+v_{\star}^2 \abs{\vb{q}}^2+\gamma^2)}{(-x^2 + v_{\star}^2 \abs{\vb{q}}^2+\gamma^2)^2} & \frac{-(-x^2+\gamma^2) v_{\star} \abs{\vb{k}} \eta e^{i \eta s_{\alpha} \theta_{\vb{k}}}}{(-x^2 + v_{\star}^2 \abs{\vb{q}}^2+\gamma^2)^2} \\[1mm]
        \frac{-(-x^2+\gamma^2) v_{\star} \abs{\vb{k}} \eta e^{-i \eta s_{\alpha} \theta_{\vb{k}}}}{(-x^2 + v_{\star}^2 \abs{\vb{q}}^2+\gamma^2)^2} & \frac{-zv_{\star}^2 \abs{\vb{q}}^2(x^2+v_{\star}^2 \abs{\vb{q}}^2+\gamma^2)}{(v_{\star}^2 \abs{\vb{q}}^2+\gamma^2) (-x^2 + v_{\star}^2 \abs{\vb{q}}^2+\gamma^2)^2}
    \end{bmatrix}, \\
    \int_0^{2\pi} \frac{d\theta_{\vb{q}}}{2\pi}\delta G^{(fc)\eta} &= 
    \begin{bmatrix}
        \frac{-2 xz\gamma  }{\left(-x^2+v_{\star}^2 \abs{\vb{q}}^2+\gamma ^2\right)^2}, & \frac{ x\gamma \eta v_{\star}(\eta k_x+is_{\alpha} k_y) (x^2 \gamma^2 + v_{\star}^4 \abs{\vb{q}}^4 - \gamma ^4)  }{\left(v_{\star}^2 \abs{\vb{q}}^2+\gamma ^2\right)^2\left(-x^2+v_{\star}^2 \abs{\vb{q}}^2+\gamma ^2\right)^2}
    \end{bmatrix}.
\end{align}
As a result, we have $\int'_q \delta G$ as follows
\begin{align}
     \int_{\Lambda-d\Lambda}^{\Lambda} \frac{\abs{\vb{q}}d\abs{\vb{q}}}{2\pi} \int_{-\infty}^{\infty} \frac{d \nu}{2\pi}\int_0^{2\pi} \frac{d\theta_{\vb{q}}}{2\pi} \delta G^{(f)} &= 0, \\
     \int_{\Lambda-d\Lambda}^{\Lambda} \frac{\abs{\vb{q}}d\abs{\vb{q}}}{2\pi} \int_{-\infty}^{\infty} \frac{d \nu}{2\pi}\int_0^{2\pi} \frac{d\theta_{\vb{q}}}{2\pi} \delta G^{(c)\eta} &= -\frac{\Lambda d\Lambda}{8\pi} \frac{(v_{\star}^2\Lambda^2+2\gamma^2) }{( v_{\star}^2 \Lambda^2+\gamma^2)^{3/2}}
     \begin{bmatrix}
        0 &  v_{\star} \abs{\vb{k}} \eta e^{i \eta s_{\alpha} \theta_{\vb{k}}} \\
        v_{\star} \abs{\vb{k}} \eta e^{-i \eta s_{\alpha} \theta_{\vb{k}}} & 0
    \end{bmatrix}, \\
    \int_{\Lambda-d\Lambda}^{\Lambda} \frac{\abs{\vb{q}}d\abs{\vb{q}}}{2\pi} \int_{-\infty}^{\infty} \frac{d \nu}{2\pi}\int_0^{2\pi} \frac{d\theta_{\vb{q}}}{2\pi} \delta G^{(fc)\eta} &= [0, 0],
\end{align}
which leads to the exchange integral
\begin{align}
    \int'_q G_1^{\alpha \eta}(q+k) &\approx \int_{\Lambda-d\Lambda}^{\Lambda} \frac{\abs{\vb{q}}d\abs{\vb{q}}}{2\pi} \int_{-\infty}^{\infty} \frac{d \nu}{2\pi}\int_0^{2\pi} \frac{d\theta_{\vb{q}}}{2\pi} [G_1^{\alpha \eta}(q) + \delta G] \nonumber \\
    &= -\frac{\Lambda d\Lambda}{4\pi \sqrt{v_{\star}^2 \Lambda^2 + \gamma^2}} 
    \begin{bmatrix}
        0 & \gamma & 0 \\
        \gamma & 0 & 0 \\
        0 & 0 & 0
    \end{bmatrix} 
    -\frac{\Lambda d\Lambda}{8\pi} \frac{(v_{\star}^2\Lambda^2+2\gamma^2) }{( v_{\star}^2 \Lambda^2+\gamma^2)^{3/2}}
     \begin{bmatrix}
        0 & 0 & 0 \\
        0 & 0 &  v_{\star} \abs{\vb{k}} \eta e^{i \eta s_{\alpha} \theta_{\vb{k}}} \\
        0 &v_{\star} \abs{\vb{k}} \eta e^{-i \eta s_{\alpha} \theta_{\vb{k}}} & 0
    \end{bmatrix}.
\end{align}

\section{RG Equations beyond the leading order}
\label{sec:RG_beyond_leading}
\begin{figure}[b]
    \centering
    \includegraphics[width=\linewidth]{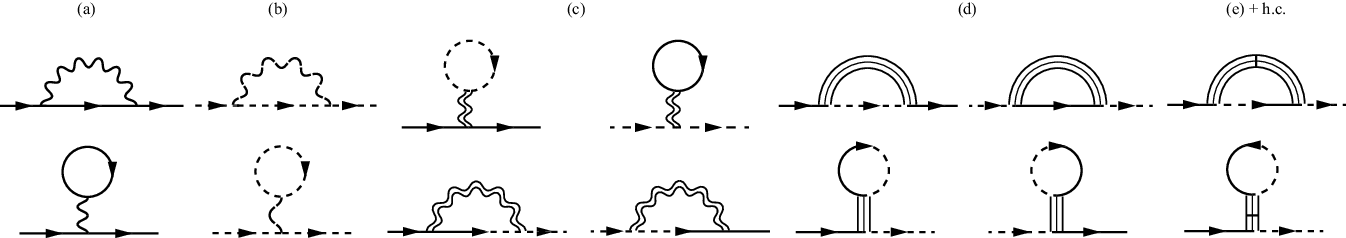}
    \caption{Renormalization to the self energy. }
    \label{fig:self_energy}
\end{figure}

\begin{figure}
    \centering
    \includegraphics[width=0.6\linewidth]{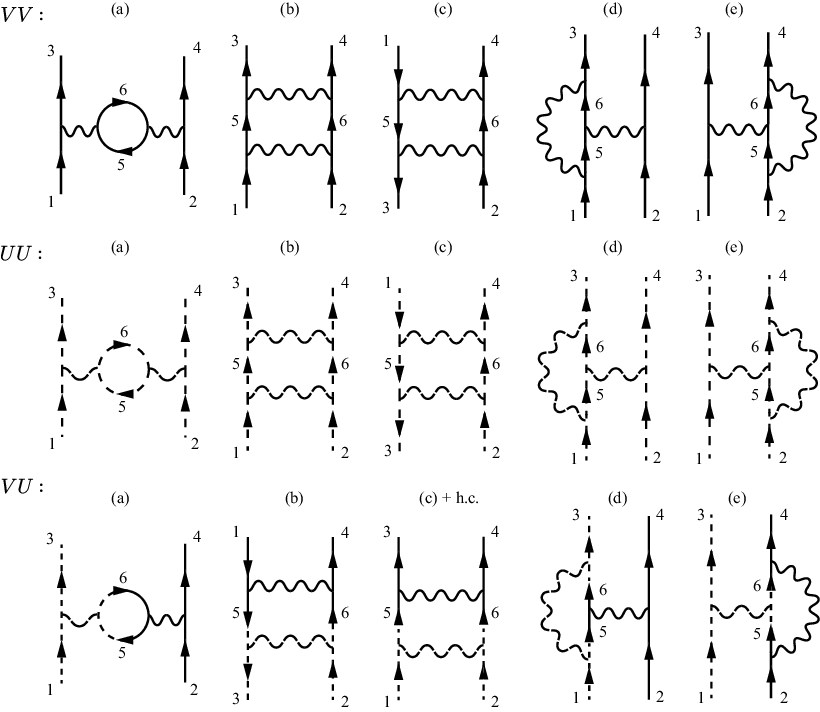}
    \caption{Renormalization to the interaction terms contributed from $VV$, $UU$, and $VU$. }
    \label{fig:VV_UU_VU}
\end{figure}

\begin{figure}
    \centering
    \includegraphics[width=\linewidth]{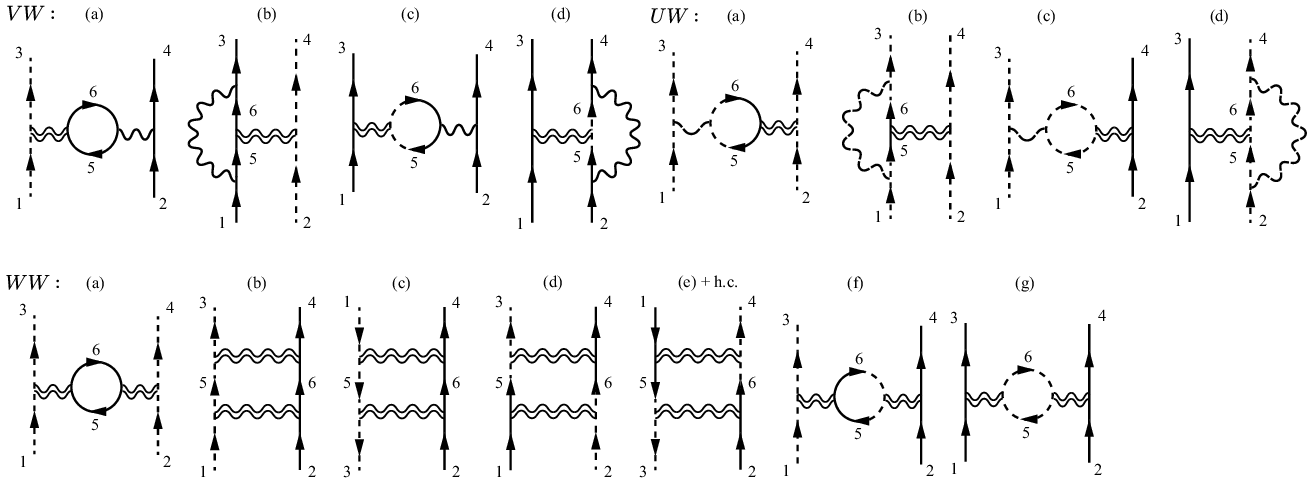}
    \caption{Renormalization to the interaction terms contributed from $VW$, $UW$, and $WW$. }
    \label{fig:VW_UW_WW}
\end{figure}

\begin{figure}
    \centering
    \includegraphics[width=\linewidth]{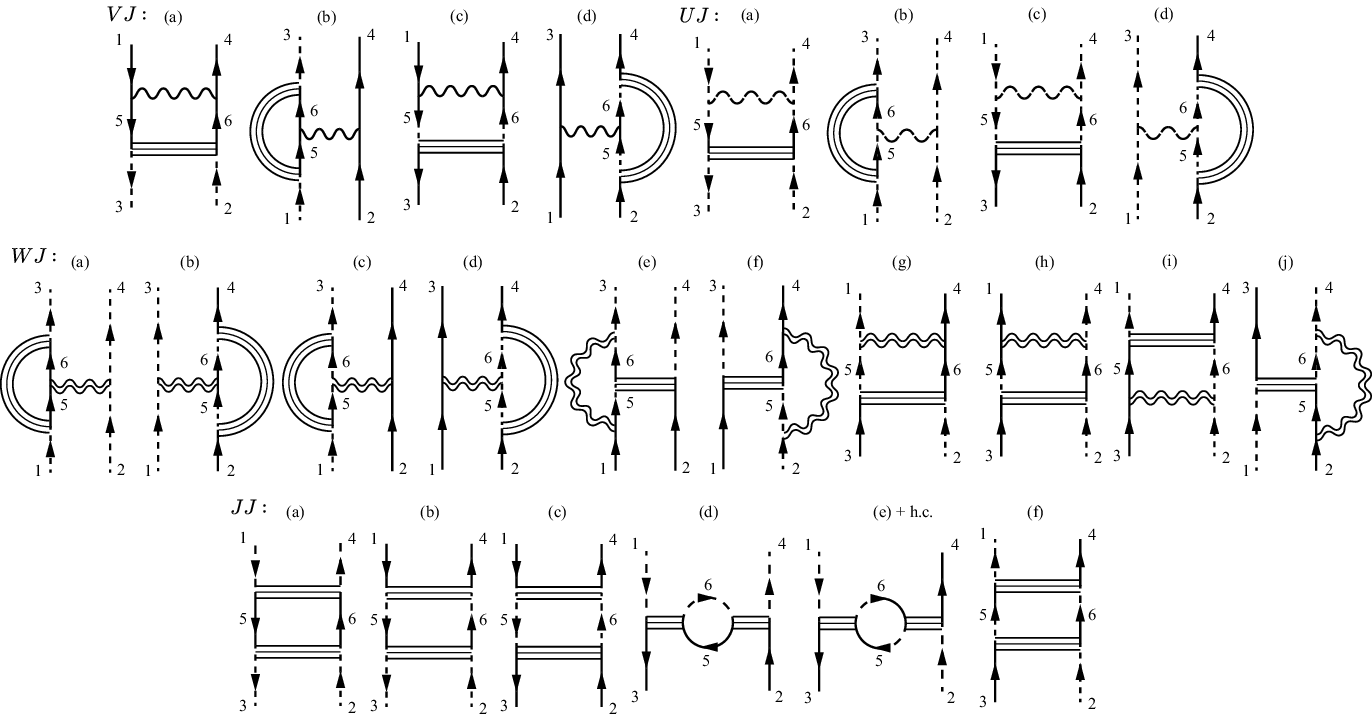}
    \caption{Renormalization to the interaction terms contributed from $VJ$, $UJ$, $WJ$, and $JJ$. }
    \label{fig:VJ_UJ_WJ_JJ}
\end{figure}

\begin{figure}
    \centering
    \includegraphics[width=\linewidth]{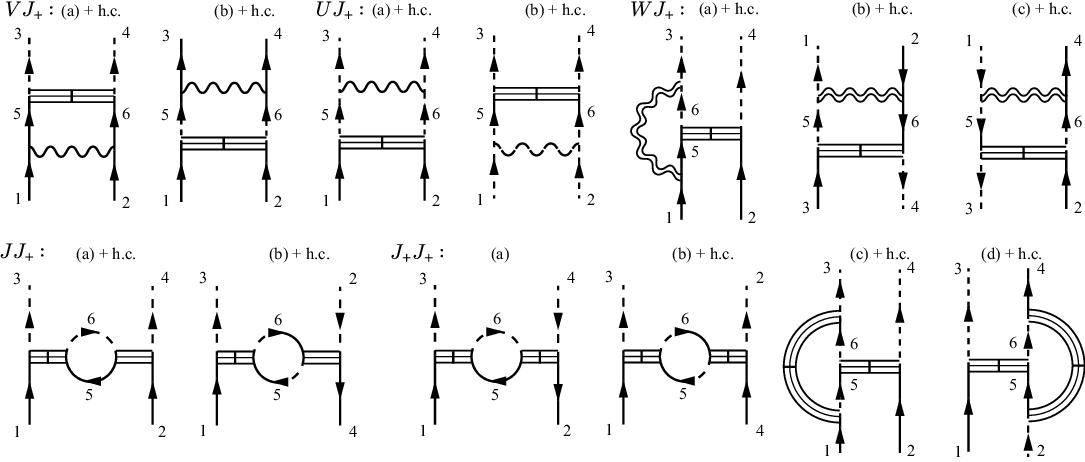}
    \caption{Renormalization to the interaction terms contributed from $VJ_{+}$, $UJ_{+}$, $WJ_{+}$, $JJ_{+}$, and $J_{+}J_{+}$. }
    \label{fig:VJ+_UJ+_WJ+_J+J+}
\end{figure}

In this appendix, we provide all one-loop diagrams including subleading terms of order $\mathcal{O}(\gamma^2/v_{\star}^2\Lambda^2)$ and $\mathcal{O}(\gamma^4/v_{\star}^4\Lambda^4)$ for completeness.
The first-order terms $\delta S_I^{(1)} = \ev{S_I}$ correspond to the self-energy shown in Fig.~\ref{fig:self_energy}.
As mentioned in the maintext, only the exchange diagrams in Fig.~\ref{fig:self_energy} (a) and (c) have non-zero contribution to the self energy.
The second-order term $\delta S_I^{(2)} = - \frac{1}{2} (\ev{S_I^2}- \ev{S_I}^2)$ corresponds to the diagrams shown in Figs.~\ref{fig:VV_UU_VU}-\ref{fig:VJ+_UJ+_WJ+_J+J+}.
To summarize, the one-loop RG contributions to the interactions are given by
\begin{align}
    \frac{dV}{d\Lambda} &= \qty[\qty(1 - \frac{1}{8}) V^2 - V(W_1 + W_3) + \frac{1}{8} V (W_1 + W_3) + \frac{1}{4} (W_1 + W_3)^2 + \frac{1}{8} VJ - \frac{1}{32} J (W_1 + W_3)] \frac{\gamma^4 \Lambda}{\pi (v_{\star}^2 \Lambda^2 + \gamma^2)^{5/2}}, \\
     \frac{d U}{d\Lambda} &= \qty[(W_1 - W_3)^2 + \frac{1}{4} J(W_1 - W_3)] \frac{v_{\star}^4 \Lambda^5}{\pi (v_{\star}^2 \Lambda^2 + \gamma^2)^{5/2}} + \nonumber \\
     &+ \qty[-\qty(2 - \frac{1}{4}) U(W_1 -W_3) + 2 W_1(W_1 - W_3) - \frac{1}{4}UJ + \frac{1}{8} J W_1] \frac{\gamma^2 v_{\star}^2 \Lambda^3 }{\pi (v_{\star}^2 \Lambda^2 + \gamma^2)^{5/2}} + \nonumber \\
     &+ \qty[\qty(1 - \frac{1}{4}) U^2 - \qty(2 - \frac{1}{4}) U W_1 + W_1^2] \frac{\gamma^4 \Lambda}{\pi (v_{\star}^2 \Lambda^2 + \gamma^2)^{5/2}}, \\
     \frac{d W_1}{d\Lambda} &= -\frac{1}{8} V(W_1 - W_3) \frac{v_{\star}^4 \Lambda^5}{\pi (v_{\star}^2 \Lambda^2 + \gamma^2)^{5/2}} + \nonumber \\
     &+ \qty[\frac{1}{8} VU + \qty(1 - \frac{1}{8}) V (W_1 - W_3) - \frac{1}{8} V W_1 - W_1(W_1 - W_3) + \frac{1}{8} VJ - \frac{1}{8} W_1 J] \frac{\gamma^2 v_{\star}^2 \Lambda^3}{\pi (v_{\star}^2 \Lambda^2 + \gamma^2)^{5/2}} + \nonumber \\
     &+ \qty[-\qty(1 - \frac{1}{8} - \frac{1}{8}) VU + \qty(1 - \frac{1}{8} - \frac{1}{8}) VW_1 + \qty(1 - \frac{1}{8}) UW_1 - W_1^2] \frac{\gamma^4 \Lambda}{\pi (v_{\star}^2 \Lambda^2 + \gamma^2)^{5/2}}, \\
     \frac{d W_3}{d\Lambda} &= \frac{1}{8} V(W_1 - W_3) \frac{v_{\star}^4 \Lambda^5}{\pi (v_{\star}^2 \Lambda^2 + \gamma^2)^{5/2}} + \nonumber \\
     &+ \qty[-\frac{1}{8} VU + V (W_1 - W_3) + \frac{1}{8} V W_1 - W_3(W_1 - W_3) + \frac{1}{8} VJ + \frac{1}{8} (W_1 - W_3) J - \frac{1}{8} W_3 J] \frac{\gamma^2 v_{\star}^2 \Lambda^3}{\pi (v_{\star}^2 \Lambda^2 + \gamma^2)^{5/2}} + \nonumber \\
     &+ \qty[-\qty(1 - \frac{1}{8}) VU + VW_1 + \qty(1 - \frac{1}{8}) UW_1 - W_3 W_1 - \frac{1}{8} UJ + \frac{1}{8} W_1 J] \frac{\gamma^4 \Lambda}{\pi (v_{\star}^2 \Lambda^2 + \gamma^2)^{5/2}}, \\
     \frac{d J}{d\Lambda} &= -\frac{1}{8} VJ \frac{v_{\star}^4 \Lambda^5}{\pi (v_{\star}^2 \Lambda^2 + \gamma^2)^{5/2}} + \nonumber \\
     &+ \qty[-\frac{1}{8} VU - \frac{1}{8} W_3^2 - (\frac{1}{8}-\frac{1}{8}+\frac{1}{8}-\frac{1}{8}) W_3 J - (\frac{1}{8} - \frac{1}{2} - \frac{1}{8}) J^2 - \frac{1}{2} J_{+}^2] \frac{\gamma^2 v_{\star}^2 \Lambda^3}{\pi (v_{\star}^2 \Lambda^2 + \gamma^2)^{5/2}} + \nonumber \\
     &-  \frac{1}{8} U J \frac{\gamma^4 \Lambda}{\pi (v_{\star}^2 \Lambda^2 + \gamma^2)^{5/2}}, \\
     \frac{d J_{+}}{d\Lambda} &= \frac{1}{8} VJ_{+} \frac{v_{\star}^4 \Lambda^5}{\pi (v_{\star}^2 \Lambda^2 + \gamma^2)^{5/2}} + \nonumber \\
     &+ \qty[\frac{1}{8} VU + \frac{1}{8} W_3^2 - (\frac{1}{4} + \frac{1}{4}) W_3 J_{+} + JJ_{+}] \frac{\gamma^2 v_{\star}^2 \Lambda^3}{\pi (v_{\star}^2 \Lambda^2 + \gamma^2)^{5/2}} + \nonumber \\
     &+  \frac{1}{8} U J_{+} \frac{\gamma^4 \Lambda}{\pi (v_{\star}^2 \Lambda^2 + \gamma^2)^{5/2}}.
\end{align}
We solve the RG equations using Mathematica and present the solution in Figs.~\ref{fig:gamma_vstar_RG} and~\ref{fig:interaction_flow}. The source code can be found in Ref.~\cite{matbg_thf_rg_2025}.
The RG contributions corresponding to diagrams shown in Fig.~\ref{fig:VV_UU_VU} read
\begin{align}
    &\text{$VV$ (a): } \delta V = - V^2 \frac{1}{\pi} \Lambda d\Lambda \frac{\gamma^4}{(v_{\star}^2 \Lambda^2 + \gamma^2)^{5/2}}, \\
    &\text{$VV$ (b)+(c)=0, } \\
    &\text{$VV$ (d)=(e): } \delta V =  V^2 \frac{1}{16\pi} \Lambda d\Lambda \frac{\gamma^4}{(v_{\star}^2 \Lambda^2 + \gamma^2)^{5/2}}, \\
    &\text{$UU$ (a): } \delta U = - U^2 \frac{1}{\pi} \Lambda d\Lambda \frac{\gamma^4}{(v_{\star}^2 \Lambda^2 + \gamma^2)^{5/2}}, \\
    &\text{$UU$ (b)+(c)=0, } \\
    &\text{$UU$ (d)=(e): } \delta U =  U^2 \frac{1}{8\pi} \Lambda d\Lambda \frac{\gamma^4}{(v_{\star}^2 \Lambda^2 + \gamma^2)^{5/2}}, \\
    &\text{$VU$ (a): } \delta W_1 = \delta W_3 = VU \frac{1}{\pi} \Lambda d\Lambda \frac{\gamma^4}{(v_{\star}^2 \Lambda^2 + \gamma^2)^{5/2}}, \\
    &\text{$VU$ (b): } \delta J = VU \frac{1}{8\pi} \Lambda d\Lambda \frac{v_{\star}^2 \Lambda^2 \gamma^2}{(v_{\star}^2 \Lambda^2 + \gamma^2)^{5/2}}, \\
    &\text{$VU$ (c) + h.c.: } \delta J_{+} = -VU \frac{1}{8\pi} \Lambda d\Lambda \frac{v_{\star}^2 \Lambda^2 \gamma^2}{(v_{\star}^2 \Lambda^2 + \gamma^2)^{5/2}}, \\
    &\text{$VU$ (d): } \delta W_1 = \delta W_3 = -VU \frac{1}{8\pi} \Lambda d\Lambda \frac{\gamma^4}{(v_{\star}^2 \Lambda^2 + \gamma^2)^{5/2}}, \\
    &\text{$VU$ (e): } \delta W_1 = -VU \frac{1}{8\pi} \Lambda d\Lambda \frac{\gamma^2}{(v_{\star}^2 \Lambda^2 + \gamma^2)^{3/2}}, \\
    &\;\;\;\;\;\;\;\;\;\;\delta W_3 = VU \frac{1}{8\pi} \Lambda d\Lambda \frac{v_{\star}^2 \Lambda^2 \gamma^2}{(v_{\star}^2 \Lambda^2 + \gamma^2)^{5/2}}.
\end{align}
The RG contributions corresponding to diagrams shown in Fig.~\ref{fig:VW_UW_WW} read
\begin{align}
    &\text{$VW$ (a): } \delta W_1 = \delta W_3 = -V \frac{1}{\pi} \Lambda d\Lambda \frac{\gamma^2[v_{\star}^2 \Lambda^2 (W_1 - W_3) + W_1 \gamma^2]}{(v_{\star}^2 \Lambda^2 + \gamma^2)^{5/2}}, \\
    &\text{$VW$ (b): } \delta W_1 = V \frac{1}{8\pi} \Lambda d\Lambda \frac{v_{\star}^2 \Lambda^2 (W_1 - W_3) + W_1 \gamma^2}{(v_{\star}^2 \Lambda^2 + \gamma^2)^{3/2}}, \\
    &\;\;\;\;\;\;\;\;\;\;\;\;\delta W_3 = -V \frac{1}{8\pi} \Lambda d\Lambda \frac{v_{\star}^2 \Lambda^2[v_{\star}^2 \Lambda^2 (W_1 - W_3) + W_1 \gamma^2]}{(v_{\star}^2 \Lambda^2 + \gamma^2)^{5/2}}, \\
    &\text{$VW$ (c): } \delta V = V (W_1 + W_3) \frac{1}{\pi} \Lambda d\Lambda  \frac{\gamma^4}{(v_{\star}^2 \Lambda^2 + \gamma^2)^{5/2}}, \\
    &\text{$VW$ (d): } \delta V = -V (W_1 + W_3) \frac{1}{8\pi} \Lambda d\Lambda \frac{\gamma^4}{(v_{\star}^2 \Lambda^2 + \gamma^2)^{5/2}}, \\
    &\text{$UW$ (a): } \delta U = U \frac{2}{\pi} \Lambda d\Lambda \frac{\gamma^2[v_{\star}^2 \Lambda^2 (W_1 - W_3) + W_1 \gamma^2]}{(v_{\star}^2 \Lambda^2 + \gamma^2)^{5/2}}, \\
    &\text{$UW$ (b): } \delta U = -U \frac{1}{4\pi} \Lambda d\Lambda \frac{\gamma^2[v_{\star}^2 \Lambda^2 (W_1 - W_3) + W_1 \gamma^2]}{(v_{\star}^2 \Lambda^2 + \gamma^2)^{5/2}}, \\
    &\text{$UW$ (c): } \delta W_a = -U W_a \frac{1}{\pi} \Lambda d\Lambda \frac{\gamma^4}{(v_{\star}^2 \Lambda^2 + \gamma^2)^{5/2}}, \\
    &\text{$UW$ (d): } \delta W_a = U W_a \frac{1}{8\pi} \Lambda d\Lambda \frac{\gamma^4}{(v_{\star}^2 \Lambda^2 + \gamma^2)^{5/2}}, \\
    &\text{$WW$ (a): } \delta U = - \frac{1}{\pi} \Lambda d\Lambda \frac{[v_{\star}^2 \Lambda^2 (W_1 - W_3) + W_1 \gamma^2]^2}{(v_{\star}^2 \Lambda^2 + \gamma^2)^{5/2}}, \\
    &\text{$WW$ (b) + (c) = 0, } \\
    &\text{$WW$ (d): } \delta J = W_3^2 \frac{1}{8\pi} \Lambda d\Lambda \frac{\gamma^2 v_{\star}^2 \Lambda^2}{(v_{\star}^2 \Lambda^2 + \gamma^2)^{5/2}}, \\
    &\text{$WW$ (e) + h.c.: } \delta J_{+} = -W_3^2 \frac{1}{8\pi} \Lambda d\Lambda \frac{\gamma^2 v_{\star}^2 \Lambda^2}{(v_{\star}^2 \Lambda^2 + \gamma^2)^{5/2}}, \\
    &\text{$WW$ (f): } \delta W_a = W_a \frac{1}{\pi} \Lambda d\Lambda \frac{\gamma^2[v_{\star}^2 \Lambda^2 (W_1 - W_3) + W_1 \gamma^2]}{(v_{\star}^2 \Lambda^2 + \gamma^2)^{5/2}}, \\
    &\text{$WW$ (g): } \delta V = - \frac{1}{4}(W_1 + W_3)^2 \frac{1}{\pi} \Lambda d\Lambda \frac{\gamma^4}{(v_{\star}^2 \Lambda^2 + \gamma^2)^{5/2}}.
\end{align}
The RG contributions corresponding to diagrams shown in Fig.~\ref{fig:VJ_UJ_WJ_JJ} read
\begin{align}
    &\text{$VJ$ (a): } \delta J = VJ \frac{1}{8\pi} \Lambda d\Lambda \frac{v_{\star}^4 \Lambda^4}{(v_{\star}^2 \Lambda^2 + \gamma^2)^{5/2}}, \\
    &\text{$VJ$ (b): } \delta W_1 = \delta W_3 = - VJ \frac{1}{8\pi} \Lambda d\Lambda \frac{v_{\star}^2 \Lambda^2 \gamma^2}{(v_{\star}^2 \Lambda^2 + \gamma^2)^{5/2}}, \\
    &\text{$VJ$ (c): New interaction} \\
    &\text{$VJ$ (d): } \delta V = - VJ \frac{1}{8\pi} \Lambda d\Lambda \frac{\gamma^4}{(v_{\star}^2 \Lambda^2 + \gamma^2)^{5/2}}, \\
    &\text{$UJ$ (a): New interaction} \\
    &\text{$UJ$ (b): } \delta U = UJ \frac{1}{4\pi} \Lambda d\Lambda \frac{v_{\star}^2 \Lambda^2 \gamma^2}{(v_{\star}^2 \Lambda^2 + \gamma^2)^{5/2}}, \\
    &\text{$UJ$ (c): } \delta J = UJ \frac{1}{8\pi} \Lambda d\Lambda \frac{\gamma^4}{(v_{\star}^2 \Lambda^2 + \gamma^2)^{5/2}}, \\
    &\text{$UJ$ (d): } \delta W_3 = UJ \frac{1}{8\pi} \Lambda d\Lambda \frac{\gamma^4}{(v_{\star}^2 \Lambda^2 + \gamma^2)^{5/2}}, \\
    &\text{$WJ$ (a): } \delta U = - J \frac{1}{4\pi} \Lambda d\Lambda \frac{v_{\star}^2 \Lambda^2[v_{\star}^2 \Lambda^2 (W_1 - W_3) + W_1 \gamma^2]}{(v_{\star}^2 \Lambda^2 + \gamma^2)^{5/2}}, \\
    &\text{$WJ$ (b): } \delta W_3 = -J \frac{1}{8\pi} \Lambda d\Lambda \frac{\gamma^2[v_{\star}^2 \Lambda^2 (W_1 - W_3) + W_1 \gamma^2]}{(v_{\star}^2 \Lambda^2 + \gamma^2)^{5/2}}, \\ 
    &\text{$WJ$ (c): } \delta W_a = W_a J \frac{1}{8\pi} \Lambda d\Lambda \frac{\gamma^2 v_{\star}^2 \Lambda^2}{(v_{\star}^2 \Lambda^2 + \gamma^2)^{5/2}}, \\
    &\text{$WJ$ (d): } \delta V = \frac{1}{2}(W_1 + W_3) J \frac{1}{16\pi} \Lambda d\Lambda \frac{\gamma^4}{(v_{\star}^2 \Lambda^2 + \gamma^2)^{5/2}}, \\
    &\text{$WJ$ (e) = 0, }\\
    &\text{$WJ$ (f): } \delta J = W_3 J \frac{1}{8\pi} \Lambda d\Lambda \frac{v_{\star}^2 \Lambda^2 \gamma^2}{(v_{\star}^2 \Lambda^2 + \gamma^2)^{5/2}}, \\
    &\text{$WJ$ (g): } \delta J = -W_3 J \frac{1}{8\pi} \Lambda d\Lambda \frac{v_{\star}^2 \Lambda^2 \gamma^2}{(v_{\star}^2 \Lambda^2 + \gamma^2)^{5/2}},\\
    &\text{$WJ$ (h): New interactions,}\\
    &\text{$WJ$ (i): } \delta J = -W_3 J \frac{1}{8\pi} \Lambda d\Lambda \frac{v_{\star}^2 \Lambda^2 \gamma^2}{(v_{\star}^2 \Lambda^2 + \gamma^2)^{5/2}}, \\
    &\text{$WJ$ (j): } \delta J = W_3 J \frac{1}{8\pi} \Lambda d\Lambda \frac{v_{\star}^2 \Lambda^2 \gamma^2}{(v_{\star}^2 \Lambda^2 + \gamma^2)^{5/2}}, \\
    &\text{$JJ$ (a), (c): New interactions,}  \\
    &\text{$JJ$ (b): } \delta J = J^2 \frac{1}{8\pi} \Lambda d\Lambda \frac{v_{\star}^2 \Lambda^2 \gamma^2}{(v_{\star}^2 \Lambda^2 + \gamma^2)^{5/2}}, \\
    &\text{$JJ$ (d): } \delta J = -J^2 \frac{1}{2\pi} \Lambda d\Lambda \frac{v_{\star}^2 \Lambda^2 \gamma^2}{(v_{\star}^2 \Lambda^2 + \gamma^2)^{5/2}}, \\
    &\text{$JJ$ (e)=0,} \\
    &\text{$JJ$ (f): } \delta J = -J^2 \frac{1}{8\pi} \Lambda d\Lambda \frac{v_{\star}^2 \Lambda^2 \gamma^2}{(v_{\star}^2 \Lambda^2 + \gamma^2)^{5/2}}.
\end{align}
The RG contributions corresponding to diagrams shown in Fig.~\ref{fig:VJ+_UJ+_WJ+_J+J+} read
\begin{align}
    &\text{$VJ_{+}$ (a) + h.c.: } \delta J_{+} = - VJ_{+} \frac{1}{8\pi} \Lambda d\Lambda \frac{v_{\star}^4 \Lambda^4}{(v_{\star}^2 \Lambda^2 + \gamma^2)^{5/2}}, \\
    &\text{$VJ_{+}$ (b) + h.c.: New interaction,} \\
    &\text{$UJ_{+}$ (a) + h.c.: } \delta J_{+} = - UJ_{+} \frac{1}{8\pi} \Lambda d\Lambda \frac{\gamma^4}{(v_{\star}^2 \Lambda^2 + \gamma^2)^{5/2}}, \\
    &\text{$UJ_{+}$ (b) + h.c.: New interaction,} \\
    &\text{$WJ_{+}$ (a) + h.c.: } \delta J_{+} = W_3 J_{+} \frac{1}{4\pi} \Lambda d\Lambda \frac{v_{\star}^2 \Lambda^2 \gamma^2}{(v_{\star}^2 \Lambda^2 + \gamma^2)^{5/2}}, \\
    &\text{$WJ_{+}$ (b) + h.c.: } \delta J_{+} = W_3 J_{+} \frac{1}{4\pi} \Lambda d\Lambda \frac{v_{\star}^2 \Lambda^2 \gamma^2}{(v_{\star}^2 \Lambda^2 + \gamma^2)^{5/2}}, \\
    &\text{$WJ_{+}$ (c) + h.c.: New interactions,} \\
    &\text{$JJ_{+}$ (a) + h.c.: } \delta J_{+} = -J J_{+} \frac{1}{\pi} \Lambda d\Lambda \frac{v_{\star}^2 \Lambda^2 \gamma^2}{(v_{\star}^2 \Lambda^2 + \gamma^2)^{5/2}}, \\
    &\text{$JJ_{+}$ (b)=0, } \\
    &\text{$J_{+}^2$ (a) + h.c.: } \delta J = - (J_{+})^2 \frac{1}{2\pi} \Lambda d\Lambda \frac{v_{\star}^2 \Lambda^2 \gamma^2}{(v_{\star}^2 \Lambda^2 + \gamma^2)^{5/2}}, \\
    &\text{$J_{+}^2$ (b), (c), (d)=0. }
\end{align}

\twocolumngrid


\begin{thebibliography}{83}
\bibitem [{c_b()}]{c_band_cutoff}%
  \BibitemOpen
  \href@noop {} {}\bibinfo {note} {Theoretically $\Lambda$ can be taken as $+\infty$ to get a full continuum model that contains all higher remote bands, although they do not match the higher remote bands in the BM model.}\BibitemShut {Stop}%
\bibitem [{mas()}]{mass}%
  \BibitemOpen
  \href@noop {} {}\bibinfo {note} {Adding small bandwidth $M$, which is a matrix element between $c_3$ and $c_4$ fermions, breaks such symmetry and hybridizes $C=\pm1$ Chern bands.}\BibitemShut {Stop}%
\end{thebibliography}

%

\end{document}